# Helium Clusters' Capture of Heliophobes, Strong Depletion and Spin dependent Pick-up Statistics

**Sascha Vongehr[1]**

Department of Physics and National Laboratory of Solid State Microstructures
Department of Materials Science and Engineering, Nanjing University, Nanjing
210093, P.R. China

This much revised and shortened PhD thesis contains many ideas that I could not follow up on, like self destructing beams in scattering cells, the depletion enhancing "Wittig tube", ionic seeding via beta-decay foil or Langmuir-Taylor filaments, analysis of the popular $<N> \sim \Delta N$ relation in droplet size distributions, etc. Avoiding pasting again the usual that is found in many a thesis in the He-droplet field, we focus instead on what is presented insufficiently rigorous elsewhere, like chopper selection, ionization yield curves, or certain widely employed yet wrong derivations. It is not telling much about successes (e.g. first observation of alkali clusters $A_k$ on $He_N$ with $k>3$, proof of their surface location, prediction of constant signal ratios via spin statistics) but goes mainly into the failures, as these are more interesting to those who like to explore truly new territory. Some ideas here may just need a single good insight of yours to turn them into success.

## Contents

[1] vongehr@usc.edu

# 1    Introduction

The study and use of beams of clusters of helium began in the early eighties and is still a growing field. As a tool, $He_N$, called droplets if $N \geq 1000$, serve as cryostats for the preparation and analysis of species that would otherwise, say by free beam expansion or conventional matrix isolation, be difficult to study. The droplets are also interesting by themselves. Clusters of $^4He$ are superfluid yet too small to be approximated by an infinite condensate. Mass abundance spectra of by helium clusters captured yet heliophobe species lead to intriguing problems. To grow alkali guest clusters ($A_k$) with $He_N$ hosts enables ultra low temperature isolation spectroscopy with these for cluster science important metal nano-particles. However, helium does not wet alkali atoms. It was thought impossible to grow sodium to sizes larger than the trimer $Na_{k>3}$.

We describe the first observation of clusters $A_k$ up to $k = 13$ on $He_N$. Their surface location is proven via establishing the shape of ionization yield curves. The observed mass spectra have been interpreted with two mutually exclusive models. Each involves a configuration $He_N A_k$ of considerable interest, namely on one hand highly spin polarized clusters on the surface, on the other metallically bound, ultra cold alkali clusters that may or may not reside on the surface of the helium droplet. We helped resolving the issue by deriving folded statistics of the capture, the spin dependent desorption and the evaporation of helium due to each capture event. Results are mostly exact or at least analyticity is preserved. A lot of this has by now been settled with much better experimental equipment [Bue06], though some questions seem still open.

A model applicable to usually employed cells for beam scattering and impurity pick-up is presented. It predicts steady states of cluster beams that self destruct in the cloud of atoms that a beam thereby continuously replenishes. The model is used to estimate whether the mechanism responsible can enhance depletion spectroscopic signals via a "Wittig tube". Helium droplet size distributions and their dispersion relations between average and standard distribution are also discussed.

## 1.1. Helium as a Matrix

Helium is a very mild matrix. As a noble gas, it is chemically inert. It is the chemical with the smallest liquid density. The inter-atomic distance is about 4.5Å although the atomic radius is only 31pm. Firstly, this is because the element helium (He) has the lowest dipole polarizability $\alpha = 0.205\text{Å}^3$ [Rad85], even lower than the already low ones of other noble gasses ($\alpha_{Ar} = 1.64\text{Å}^3$). Corresponding to the low $\alpha$, He-He has the smallest van der Waals attraction [Wha94, Tan03] (vdW radius: 140pm). Secondly, due to its low mass and therefore high QM zero point energy, its condensed state is only a third as dense as it would be classically.

The extraordinarily weak interactions between the atoms cause the two stable helium isotopes to have the lowest boiling points of all substances, that is 4.21K for $^4$He and 3.19K for $^3$He [Wil87]. For most species it provides the smallest perturbation of any matrix. Embedded guest particles may have slightly red shifted spectra due to the polarizability of helium (attractive part of the potential) and thus a lower rate of spontaneous emission (rate is proportional to the emission frequency cubed) or a slight blue shift coming from the collective Pauli repulsion of the helium's s-electrons. Helium has no triple point. Due to the strong zero point motion it stays in the liquid phase even at zero K. It has already condensed in momentum space and must be pressurized to 25atm for $^4$He (34atm for $^3$He) to solidify in real space. Even at very low temperatures, there is almost no inhomogeneous broadening due to the sampling of different possible matrix sites in a solid lattice. Other noble gas matrices at low temperatures trap in interstitial or substitutional sites. Yet even compared to the fluid phase of any other noble gas, helium shows little inhomogeneous broadening, because although homogeneity always breaks down close to an impurity, in helium, the surround is determined by the dominating interaction between the impurity and the helium rather than by the interaction between atoms of the matrix [Kan97].

$^4$He below $T_\lambda = 2.172$K (2.65mK for $^3$He) is super fluid, i.e. the viscosity is lowered

by 11 orders of magnitude (from $\eta$ = 3.5 mg m$^{-1}$s$^{-1}$ at 4K). Therefore, the rotational relaxation of captured species can be suppressed and rotational spectra sharply resolved. This is impossible in a classical liquid, because random collisions destroy rotational coherence and the rotational spectrum collapses into a broad band whose width gives the rotational diffusion time. Impurities move frictionless in super fluid helium and quickly find each other. The super fluid has a six orders of magnitude larger thermal conductivity than helium above the lambda point. The thermal conductivity is then a 1000 times that of room temperature (R.T.) copper (30 times if at the same temperature). Thus, released binding energy is carried away immediately. In this way, the cold helium environment can stabilize metastable states, weakly bound adducts and transients.

Helium, especially $^3$He, is sometimes referred to as the world's purest chemical since few compounds dissolve in it. This leads to the disadvantages of the use of bulk helium. It is not easy to put material into bulk helium in the first place. Once forced inside via a beam or maybe laser ablation, the materials have a propensity of wandering to the surface or the container wall, to which they stick [Sil84] because the interaction with the wall atoms is almost always stronger than that with the helium. The difficulties with the implantation of impurities and control of their concentration have limited experiments in bulk helium to atoms, ions and dimers [Tak96]. It is very hard to work under conditions ensuring a transient regime before condensation of impurities.

## 1.2.  Clusters replacing Bulk Helium

Helium is expanded supersonically resulting in a beam of droplets which capture almost anything in their way with basically their geometrical cross section. Thus, very many tiny helium samples trap species in large concentrations inside the beam while at low concentration in any one helium cluster. Unwanted condensation is avoided, thus helium clusters are ideal for isolation spectroscopy [Gou85], called "He$_N$ Droplet

Isolation" (HENDI).

Helium atoms are bound to the surface by 0.62meV. This corresponds to $k/(2\pi) = 5\text{cm}^{-1}$ or roughly 7.15K via $k = k_B T/(\hbar c)$ (2.7K for $^3$He). The average time $\tau$ for a helium atom to evaporate from helium bulk surface at temperature $T$ is given via $\ln[\tau/\tau_0] = 7.15\text{K}/T$. Bulk helium can be held at different temperatures, but for clusters in a fast beam of velocity $v$, $T$ is determined by the time of flight $\tau \approx L/v$ through an apparatus while $\tau_0$ is proportional to the cluster cross section. For helium clusters at $T = 1$K holds $\tau_0 \approx 10^{-10}$s [Bri90] corresponding to ultra low temperatures after very short times. This has been experimentally confirmed as (0.37±0.05)K [Har97] (0.15K for $^3$He [Toe04]) for $\tau \approx 1$ms.

There is thus little temperature broadening of spectroscopic lines. Rovibrationally cold spectra can be taken. $^4$He is automatically below its lambda point (2K in clusters of $N = 1000$ atoms [Ram93]) and super fluid as long as the number of particles is $N \geq 60$ [Gre98]. Condensation is guaranteed: molecules attract each other almost unhindered by the helium in between and quickly meet ($t < 10^{-8}$s) [Lew95].

Trying the same expansion procedure with other noble gasses will result in warmer yet solid clusters because their surface tension overcomes the solidification pressure. A solid matrix brings with it inhomogeneous broadening. Spectral line shifts in helium are often roughly 10% of those in argon or neon clusters.

The $^3$He cluster temperature is above its super fluid transition. It is a Fermi liquid that is chemically much like $^4$He. Alternating $^4$He and $^3$He probes the effects of super fluidity in $^4$He since $^3$He has a viscosity of $\eta = 19.5\text{mg m}^{-1}\text{s}^{-1}$ [Bet63].

A disadvantage of clustered helium is that the temperature is indeed always close to the above stated ones. An expansion of a mixture of the isotopes of helium is impractical, because they phase separate.

The evaporative cooling after pick-up of impurities is fast. The time of equilibration is smaller than $10^{-8}$s [Gsp95, Har95, Bri90]. About $10^2$ atoms evaporate per captured molecule. Annealing of guest species is quite impossible. The cooling is too fast. This stabilizes metastable states like radicals, linear chains of $(HCN)_{n\leq 8}$ (free $(HCN)_{4\leq n}$ are cyclic) [Nau99, Kof87] and cyclic $(H_2O)_{n\leq6}$ [Nau99], etc. [Hig96, Sti95, Har96, Lin99]. It enables studies of weakly bound complexes and thereby insights into intermediates and the transition state region.

That helium evaporates so readily leads to strong beam depletion. Exothermic reactions can be monitored via the disappearance of larger $He_n$ fragments from the beam or via the deficit in energy that is deposited into a bolometer. The bolometer can also show positive energy readings in case a species gets excited but does not relax.

When a guest is embedded or attached to a helium cluster, the whole cluster can capture other particles and lead to reactions between them. The reaction $Ba + N_2O \rightarrow BaO + N_2$ has been enhanced [Lug00] 3000 times over the gas phase cross section by using droplets with $<N> = 2x10^5$ atoms, thereby creating an average geometrical cross section of $<\sigma> = 5.3x10^4 \text{Å}^2$. With such cross sections, one needs only vapor pressures of a few $\mu$Torr to implant particles into the drops [Lew95]. This is particularly useful for fragile, non-volatile compounds such as nucleotide bases and amino acids and for low vapor pressure systems like inorganic salts.

Fragile components that fragment if directly bombarded with electrons can be softly ionized inside helium clusters. This is called fragmentation quenching [Fed99] or, if energies are below helium's ionization potential $IP_{He}$, Penning ionization [Sch93]. The first method starts with a helium atom being ionized (hole creation). The guest is ionized via charge transfer from $He^+$ or $He_2^+$. This gives at most $IP_{He}$ to the guest - much less than direct electron impact at usually about 70eV.

One has to be somewhat careful when interpreting cluster and bulk spectra. The emission is shortened by the index of refraction ($\tau \sim N^{-2}$) [Hir80] but the latter does not hold for clusters that are smaller than the wavelengths used.

Aggregation is not a simple matter in non-equilibrium expansions. The difficulties encountered in trying to control, characterize, and manipulate cluster formation are severe, especially when mixed clusters are desired (phase separation). Inside helium droplets, growing mixed clusters is easily accomplished via sending the droplets through mixed gasses or sequential pick-up cells if a certain sequence (core-shell structure of internally assembled guest cluster) is desired.

### 1.3. From Stagnation Conditions to Clusters

The He source employs very small pinhole nozzles with usually (5±1)μm diameter. They require a very high stagnation pressure $P_0$. Research grade helium gas at $P_0$ 10bar to 80bar is mostly used. Above this range, the flux is usually too strong to be handled by diffusion pumps. Given the short converging nozzles for strongly supersonic beams, the expansion can be considered as isentropic if no diverging section is present. The three degrees of freedom (DOF) for the source are the stagnation condition ($P_0$, $T_0$) and the nozzle size. However, any small region of helium has no knowledge of where along the isentrope the expansion started, of the beam velocity or of how big the nozzle was, so there is only one continuous DOF in the cluster size distribution.

### 1.3.1. Sub and Super Critical Expansion Regimes

Helium can mostly be regarded as an ideal gas. It is monoatomic, so the ratio of specific heats is $\gamma = C_P/C_V = 5/3$. When an ideal gas expands adiabatically, its state follows an isentropic trajectory with $P \propto T^{\gamma/(\gamma-1)}$, being linear in a double logarithmic $P$ versus $T$ plot [Buc90]. These isentropes $\ln(P/P_0) = \gamma[\ln(T/T_0)]/(\gamma-1)$ either hit the vapor-liquid coexistence line (bi-nodal) below the critical point or they pass above it and miss the line. This gives rise to the sub- and super critical regimes respectively. Super critically, linear isentropes would hit the solidification or super fluidity transition (λ-line), but strengthening particle interaction diverges the gas from the ideal model and the isentropes bend down to avoid them. They thus *also* intersect the bi-nodal, but this time approaching from the liquid side. The two regimes are separated by the critical isentrope $S_c = 22.81$J/(K mol) that goes through the critical point $(P_c, T_c) = (2.275\text{bar}, 5.2014\text{K})$ at $\rho_c = 69.64\text{g/l}$, thus $s_c = 5.70\text{kJ/(kg K)}$ [Ang77]. It can be in practice started at $(P_0, T_0) = (30\text{bar}, 10.2\text{K})$. Applying $2\ln(P/P_0) = 5\ln(T/T_0)$ at $T = T_c$ results in $P \approx 2P_c$, so the formulas and assumptions are a rough guide only; problematic are identifying the measured $T$ and $P$ of the source with the actual values at the nozzle throat and also assuming adiabaticity in the first place.

For **sub critical expansions**, the helium at the nozzle's exit is gaseous and the clusters condense in the expansion after leaving the nozzle. This is known [Lew95] to give clusters of average sizes of $<N> = 100$ particles up to $<N> = 20\text{k}$ particles with average radii of roughly $R \approx 1\text{nm}$ to $10\text{nm}$. The radii and cross sections of droplets, that means $N > 1000$, follow $\sigma_{geo} = \pi R^2$ with $R = r_s N^{1/3}$ and the liquid Wigner-Seitz radius of $r_s = 2.221\text{Å}$ ($2.44\text{Å}$ for $^3\text{He}$) [Bri90]. Clusters below a thousand atoms have a large fraction of their atoms inside the less dense surface and are not well described by this. The small droplets and clusters have the advantage of picking up less background gas and one can neglect volume excitations. All excitations are surface ripplons because the breathing modes of small droplets with $N < 10^6$ are too high in energy to be excited at $0.37\text{K}$. The same holds even for the compression modes of clusters smaller than $N = 1000$. Small helium clusters are expected to be as mild a matrix as larger ones. Even very small $^4\text{He}$ clusters have almost no shell structure [Ram90, Mah96] (Shell structure was calculated for $^3\text{He}$ clusters though [Pan86].)

A **supercritical expansion** [Buc90] yields droplets with 30k to $10^8$ particles, above which there is the Rayleigh breakup regime [Gri03] which will not concern us here. From the fluid state at the nozzle exit, supercritical expansions result in fractionation clusters showing an exponential size distribution [Jia92] for large droplets, yet they also have a small proportion of clusters that re-condensed from single atoms after the fractionation cloud expands further and thus cools further. The overall size distribution is therefore bi-modal.

### 1.3.2. Beam Velocity

In the absence of condensation, the beam's terminal velocity $v_\infty$ is due to energy conservation $\frac{1}{2}mv_\infty^2 = h_0 m$. $h_0$ is the gas enthalpy per particle at the stagnation conditions. The ideal gas enthalpy $h_0 = k_B T_0 \gamma/(\gamma-1)$, that is $h_0 = 5(k_B T_0)/2$ for helium, is used to calibrate the source temperature sensors via time of flight measurements (condensation is then avoided). The onset of condensation though implies that one has already left the domain of applicability of ideal gas laws.

Using data for $h_0$ [Ang77] instead have predictive power over a somewhat wider range of conditions but it also cannot be good for cluster beams, because it predicts a point where the velocity goes below zero, i.e. it must be modified where condensation starts to feed energy into the expansion.

In the sub-critical regime, depending on $T_0$, cluster speeds fall in the range of $<v> = 200$m/s to $500$m/s. As a result of a quantum effects in the collisions between helium atoms [Tan95] the speed ratios $S := <v>/\Delta v$ are very high at about $S = 40$ to $100$, considering that such speed ratios are usually only attained if there is no clustering involved. The speed ratio $S$ obeys $k_B T_{\shortparallel} S^2 = \frac{1}{2}mv_\infty^2$ with $T_{\shortparallel}$ being the parallel temperature of the beam. For mono-atomic species at $S > 10$ one approximates [Toe77] using FWHM($v$) $\approx 1.65\Delta v$ and $2T_{\shortparallel}S^2 \approx 5T_0$.

A comprehensive study [Buc90] of the time of flight spectra of sub- to (including) supercritical expansions revealed that a single source condition can result in up to three differently paced beam components. The large droplets from the fractionation of the fluid helium of supercritical expansions can be quite slow close to the critical point, and speeds below $50$m/s have been observed [Haŕ97]. This results in very low speed ratios because of the bimodal size distribution; $<v>/\Delta v < 3.2$ were reported [Haŕ97]. This may be due also to unstable source conditions for example due to large density fluctuations given that the correlation length diverges at the critical point. Also, a cryostat has often plenty of cooling power above the temperature for supercritical expansions, cooling down through that temperature rapidly. Below it, the helium flow becomes very large and the cryostat may be working all of a sudden at conditions exceeding its limits on cooling power. This (or similar scenarios localized at the nozzle) can lead to a rapid oscillation in between expansion regimes. This is why our setup mostly worked near critical expansion, where the speed ratios are lowest. The cryostat exhausted its cooling power inside the transition region. All power was choked by the latent heat of the vapor to fluid transition and the onset of a much stronger mass flow.

### 1.3.3. Size Distributions

Some formulas depend on parameters that drastically change if the expansion were to occur into a plane or a higher dimensional space or if the shape constraints are of a different number of dimensions (e.g. long slit apertures instead of circular nozzles). There are some 2D and 3D molecular dynamics simulations (of super critical beams) [Ash99], that had expansion parameters $\eta$ restricting the outflow, e.g. $\eta_x = \eta_y = \eta_z$ (symmetric-triaxial), $\eta_x = \eta_y$ (beam expansion), or $\eta_x = \eta_y$ and $\eta_z = 0$ (biaxial plane strain). The result was only a small dependence on the dimension and a size distribution dependent just on the sum $\eta_v = \eta_x + \eta_y + \eta_z$.

In the **sub-critical** regime, one may estimate the mean number of atoms of the clusters $<N>$ with an empirical formula [Hag87, Hag92] using the Hagena parameter $\Gamma^*$. That formula, $<N> = 33(\Gamma^*/1000)^{2.35}$; agrees with experiments if $\Gamma^* < 7500$ [Mon01]. The subcritical regime has been suggested [Dor03] to need a further division. For argon in the range $10k < \Gamma^* < 10^6$, the formula $<N> = 100(\Gamma^*/1000)^{1.8}$ is found instead. The formulas' intersect at $\Gamma^* = 7500$, but the authors [Dor03] do not validate by going through the transition to recover Hagena's slope of $\ln<N>$ versus $\ln\Gamma^*$, thereby excluding a systematic error (one must therefore hope there was no reverse engineering of the intercept at $\Gamma^* = 7500$).

Anyways, the models are not in good agreement with today's massively clustering expansion experiments, especially not if the light rare gasses neon or helium are involved. The Hagena parameter has been recast using parameters of the fluid at the intersection of expansion isentrope and bi-nodal [Knu95, Knu97]: $\Gamma = (Kin^q)*Therm^{(1-q)}$. The kinetic term $Kin = \tau_{flow}/\tau_{nucl}$ is the ratio of a characteristic time for the matter flow ($d_0/c_0$, i.e. the nozzle diameter over velocity of sound) and a nucleation related time. As always, the subscript zero indicates conditions present inside the nozzle throat. $Therm = (P_0/A)*(T_{ref}/T_0)^{\gamma/(\gamma-1)}$, $A$ is here a vapor pressure constant, and also $T_{ref}$ is related to real fluid parameters, namely $(k_B T_{ref}) = v^{2/3} s$ where $s$ is the surface tension and $v$ the atomic volume. Most parameters can be taken from tables on bulk properties (For $^4$He: $\gamma = 5/3$, $A = 7.8$bar, $T_{ref} = 3.23$K, $v = 27.37$ml/mol). Expressions resemble closely the original Hagena results but are valid on a larger domain: $<N> = 19.138\Gamma^{-2.02}$; $1k \leq <N> \leq 10k$; $6 \leq \Gamma \leq 30$. The deviations from simpler formulas that are observed in expansions of light atomic gasses (like Ne and He) are captured by the explicit dependence on $c_0$ and $s = T_{ref} P_0/(n_0 T_0 v^{2/3})$. Previous alternatives to $\Gamma$ can be closely approached by substituting those variables with help of well known ideal gas laws: $mc^2 = \gamma(k_B T)$ and $n = P/(k_B T)$. The gasses deviating from these descriptions are of course those not well captured by the ideal gas model, especially helium.

For mono-atomic gasses ($\gamma = 5/3$) one can derive that $\Gamma$ is proportional to $P_0 d_0^q T_0^{(q-10)/4}$. The power has been fixed experimentally to $q = 4/5$.

The (near) **critical** expansion is covered at different places above. At the critical point, the surface tension vanishes and the speed of sound is minimal, etc.; all these lead to source instabilities and strongly bimodal distributions beyond the scope of this work.

In the **supercritical** regime, an exponential decay $\sim \exp[-N/<N>]$ has been observed for very large clusters and an empirical formula has been given as dependent again on surface tension $s$ and suchlike [Knu99]: $<N> = (80/3)\pi[(s/m)\tau_{flow}^2]$. Molecular dynamics calculations [Ash99] confirmed the model qualitatively. The restriction on the validity comes mainly from the fact that the super critical expansions are bimodal. It is not known whether a removal of the lognormal fraction would result in the remaining distribution being exponential also at small cluster sizes and there are reasons to doubt it [Ast00].

### 1.4. Impurities

#### 1.4.1. Ions, Electrons and Ground State Atoms

Helium clusters embed all **ions** due to the low but still existing polarizability. Positive ions attract the helium so strongly that the solidification pressure is overcome and a so called snowball of about 6Å radius develops [Atk59]. The effective mass of such a snowball with the ion kernel is roughly $m_{eff} \approx 50\ m_{He}$ [Sch75].

An **electron** can only be attached to the surface if the clusters have above $N = 5*10^5$ atoms [Ram88, Ros94, Roś94] - else the electron is not bound. Up to $N = 10^7$ there exists only one bound surface state. Inside helium, an electron strongly Pauli repulses the s-electrons of the helium atoms and a bubble ("bubblon") forms around it. The bubblon radius is about 17Å in $^4$He [Poi72, Poi74] (20Å for $^3$He [Aho78]). Given about 20eV, an electron may (if it does not excite a helium atom) penetrate over 23Å [Anc94, Anc95] into the droplet so that the then forming bubblon does not burst immediately inside the surface. The surface is defined as the 90% to 10% density falloff, $\Delta R = (6.5 \pm 0.5)$Å for $^4$He and $(8.5 \pm 0.5)$Å for $^3$He [Str87]. The effective mass of the bubblon is $m_{eff} = 250\ m_{He}$ [Kha89].

This bubblon state is metastable because the electron has 0.1eV less energy outside of the cluster. The bubblon needs to be at least about 35Å into the $^4$He droplet to be somewhat stable. Thus, there is a minimum size for the latter being $N = 75*10^3$ [$R = 93$Å] experimentally [Far98].

How **atoms** bind to helium is predicted by the Ancilotto model. The Ancilotto model uses $\lambda = n\varepsilon r_e/(2^{1/6}\sigma)$ with the surface tension $\sigma = 17.9$Å$^2$/m, number density $n = 22$/nm$^3$, and $\varepsilon$ and $r_e$ are the well depth and equilibrium bond distance of the helium-impurity potential. Density functional calculations predict that species with $\lambda \geq$ 1.9 are solvated by helium [Anc95, Anć95, Anĉ95, Ler95, Dal94]. If the Pauli repulsion between an atom's and helium's s-electrons [Tab94] dominates, the atom is heliophobe. Most such atoms only bind weakly to the surface of helium in dimple-like sites. Being much heavier than electrons, their zero point energy motion is much smaller and they therefore bind even to the smallest helium clusters. A model [Ler95] for this dimple site places a sodium atom at $z = +1.64$Å above the surface of the droplet. The dimple is described by a helium free sphere with the center at $z = +0.68$ Å above the surface and a radius of $R = 4.1$Å. The dimple therefore does not extend below the 90% to 10% density fall-off. Its effective mass is half of a bubblon's mass, which in turn is only half of a bubblon's mass at bulk density (= ¼$250m_{He}$) which is further reduced now by the lower radius $R$ as compared to the 17Å for a e⁻−bubblon. Hence, its effective mass is quite small with $m_{eff}$ = ¼250 $m_{He}$ (4.1/17)³ = 0.89 $m_{He}$. If forced into helium, they develop bubblons around them. For example, Na atoms put into bulk helium develop a $r = 5$Å bubblon in about 1ps.

Hydrogen atoms are heliophobe. Their binding towards helium is only 1.04K [Bel86]. Alkalis ($\lambda \approx 0.7$) and many alkaline-earth metals are also not solvated and stay in dimple-like sites with binding energies of 1.6 to 1.9meV[Anc95]. Li, Na and K [Hig98], Rb, Cs, Ca, Sr and Ba atoms and even Mg atoms are heliophobe, but Mg atoms [Reh00] do not stay on the surface. They sink into helium in the center of their bubblons because the involved surface energy is smaller than the energy cost of being in a dimple on the helium's surface.

### 1.4.2. Clusters and Excited States

Mg has been grown to very large clusters of hundreds of atoms in helium droplets [Die01, Dop01]. While Ca and Ba dimers stay on the helium's surface, $Ca_k$ and $Ba_k$ for increasing $k$ sink into it inside their bubblons.

Weakly (VdW) bound alkali dimers and mixtures like KNa [Hig98] and medium sized Na and K clusters have been produced and shown to stay on the surface of the helium [Von02, Sch04]. The spectra of sodium atoms and weakly bound sodium dimers and trimers residing on droplets have been investigated [Sti95, Sti96].

Strongly bound alkali clusters have not yet been found on or in $He_N$ because the release of the large covalent (metallic) binding energy leads to immediate desorption from the surface. It is not known at what size such clusters sink into helium. They must sink, because bulk alkalis, with the exception of Cs, are wetted by helium.

Atoms and molecules in high Rydberg states must be heliophobe. If produced inside helium, the metastable $2^3S$ state $He^*$ will make a bubblon and rise rapidly to the surface. As with alkali clusters, the production of a complex may lead to desorption even if the binding energy to helium is large. Electronic excitation of Na in its dimple leads to the building and desorption of a $Na^*He$ exciplex [Sti95 ,Sti96] with a binding energy of $500cm^{-1}$ - much larger than the He-Na binding.

## 1.5. Doping/Pick-up and Simple Depletion

When helium clusters $He_N$ (the hosts) travel through so called "pick-up" or "scattering" cells, they capture say $k$ guest atoms or molecules. The probability not to pick up any guest is $P_0 = e^{-<k>}$ and depends on $<k> := nFL\sigma$, where $L$ is the cell's effective scattering path length, $n$ the dopant vapor's particle number density and $\sigma$ the clusters capture cross section, which is for large helium clusters basically the geometrical cross section. $F$ involves the Maxwell-Boltzmann velocity distribution of the vapor [Ber62]:

$$F_{(x)} = \frac{1}{x\sqrt{\pi}} e^{-x^2} + \left(2 + \frac{1}{x^2}\right) \frac{1}{\sqrt{\pi}} \int_0^x e^{-t^2} dt$$

The argument is the ratio $x = <v>/\hat{u}$ of droplet speed $<v>$ and the most probable speed $\hat{u}^2 = 2(k_BT)/m$ of the scattering gas.

It can usually be assumed that pick-up events are independent of each other and that the helium droplets are massive enough to pick up without being deflected. If that holds true, then $P_0$ can be used to straightforwardly deduce the Poisson distribution $P_k = <k>^k e^{-<k>}/k!$. The average number of picked up particles is then $<k>$. $P_0$ is independent of physical complications due to pick-up and thus going via $P_{k>0} = 1-P_0$ is the proper derivation. First assuming a Lambert-Beer relation for the derivative of the probability of finite $k \neq 0$ and then solving coupled differential equations is anticipating the result and improper.

The pick-up of $k$ guests becomes maximal for $P_k = P_{k-1}$. This will change due to the broad size distribution of helium droplets. Also, without considering depletion, just because of the droplet size distribution, the Poisson approximation significantly underestimates $P_k$ at high dopant vapor densities where $<k> \geq k$ [Von10]. For isolation spectroscopy, one usually desires to isolate one guest per helium droplet. Therefore, one has to assure that the vapor pressure in the pick-up cells is just enough to have on average $<k> = 0.7$ guests per droplet.

To grow clusters inside helium droplets one mostly desires to maximize $<k>$. At strong doping, one may neither neglect the depletion due to evaporation of helium atoms dissipating kinetic, binding and condensation energy of picked-up species, nor the depletion due to deflection from picked up momentum of guests and rms momentum from the evaporation. The evaporative cooling off of excess energy after each impact of a guest into the droplet introduces dependence. Helium is very easily evaporated and the host droplet almost instantly is smaller, the geometrical cross section gets smaller, and further pick-up of guests less likely. The probability of a pick-up changes on a time scale shorter than the average time between two pick-up events; the latter are therefore *dependent* and a Poisson model is improper.

A first step in dealing with concurrent depletion is to apply the Lambert-Beer law: $I_{(n)} =: I_{(0)} e^{-\kappa}$. Similar to Ohm's law $R := U/I$ and Newton's law $p := mv$, this is neither an axiom nor a derived identity but plainly the definition of $\kappa$. The beam intensities $I_{(n)}$ at several $n$ may be measured on the helium dimer signal $He_2^+$ of a mass spectrometer. Disregarding magic numbers for neutral helium clusters and the magic $He_4^+$-ion after ionization of large helium droplets, the ratios $I_{(n)}/I_{(m)}$ as measured on the dimer after electron impact ionization is representative of the ratios that apply to the total, neutral beam intensities. $\kappa$ thus models the overall signal loss due to evaporation, deflection, loss of detection cross section, etc. If one does not measure just the helium signal and if $n \neq 0$, then the total beam intensity $I_{(n)}$ is composed of all the intensities $I_{k(n)}$ of droplets with $k$ picked up guests. That is, $I_{k(n)} := I_{(n)}P_k$ with the sum of all $P_k$ being unity and $I_{(n)} = \sum_{k=0}^{\infty} I_{k(n)}$ . Therefore, the following holds true *without* approximation: $I_{k(n)} = I_{(0)}\,P_k\,\mathrm{e}^{-\kappa}$.

Assuming Poisson statistics $P_k = {<}k{>}^k\mathrm{e}^{-{<}k{>}}/k!$ at this point not only assumes the pick up events being independent of each other. It moreover silently circumvents addressing that the Lambert-Beer attenuation $\mathrm{e}^{-\kappa}$ is not at all the same for all $k$. The more the helium clusters picked up guests, the smaller is the pick-up cross section, the more they are deflected, evaporated and also lost ionization cross section in the detector, and so on.

Concentrating on the $k = 1$ monomer signal $I_{1(n)}$, the Poisson probability makes this equal to $I_{(0)}\,{<}k{>}\,\mathrm{e}^{-(\kappa+{<}k{>})}$ [Or generally: $I_{k(n)} = I_{(0)}\,({<}k{>}^k/k!)\mathrm{e}^{-(\kappa+{<}k{>})}$]. Derivation (') with respect to $n$ yields $I_{1(n)}{'} = I_{1(n)}[{<}k{>}{'}/{<}k{>} - {<}k{>}{'} - \kappa{'}]$. Thus, if larger guest clusters do not fragment into monomers at detection, varying the vapor pressure of the dopant {For general $k$: $I_{k(n)}{'} = I_{k(n)}[k({<}k{>}{'}/{<}k{>}) - {<}k{>}{'} - \kappa{'}]$}, one encounters a maximum of the monomer signal at $I_{1(n)}{'} = 0$, that is ${<}k{>}{'} = {<}k{>}\,({<}k{>}{'} + \kappa{'})$. Only if $\kappa{'}$ is zero, like for example if $\kappa = 0$, is the maximum at ${<}k{>} = 1$.

An attenuation cross section [Hes99] can be defined via $\kappa =: nFL\sigma_{\mathrm{att}}$. Together with ${<}k{>} = nFL\sigma$ and assuming that $\sigma_{\mathrm{att}}{'}$, $\sigma{'}$ and $F{'} = 0$, this yields $nFL(\sigma_{\mathrm{att}}+\sigma) = 1$. Since $\sigma_{\mathrm{att}}$ may be established with the $He_2^+$ signal for example, finding the vapor density that leads to maximum monomer pick-up results in an estimate of the pick-up cross section $\sigma$.

This method is a first correction valid at what could be thereby defined as "simple depletion". Measurements done with CsCl as dopant [Von04] showed for example $\sigma_{\mathrm{att}}$ to be strongly dependent on the dopant vapor density $n$. These measurements were done both at temperatures where the $(CsCl)_2$ vapor pressure was still negligible and where it mattered. CsCl has a very large electric dipole moment. The reaction $2(CsCl)\rightarrow(CsCl)_2$ releases 3eV of binding energy. In this case, further pick up is strongly dependent on how many molecules have been picked up already. The methods described here are insufficient to describe the processes.

The pick up of many sodium atoms turns out to be much more involved still because the binding energy depends on the spin alignment of the $k$ guests' single spins and thus cannot be captured by a simple dependence on $k$.

## 1.6.  Penning Ionization and Ionization Yield Curves

The heliophobe alkalis are difficult to dope onto helium droplets. The mass spectrometer signal of alkali clusters remains hidden behind the intense $He_n$ fragment peaks. Penning ionization detection suppresses the signal of matrix fragments. The electron energy is set below the threshold of matrix atom ionization [$IP_{He}$ = $(1.34)^2*13.6eV = 24.6eV$] plus the 1.5eV shift from energies due to the barrier an electron has to overcome to enter the matrix, bubblon production etc. Helium can then only be excited to the $2^1S$ [20.62eV], the $2^5P$ [at $(20.35 \pm 0.15)eV$ in liquid helium], and the $2^3S$ [19.82eV] states. While the P state decays in about 16ns, the S states are metastable because the ground state cannot be reached via photon emission. These metastable states (written $He^*$) have gas phase lifetimes of 8ks. Ionization of impurities with low ionization potential ensues via collision with the metastable $He^*$. This is called Penning ionization. As a result, the helium ion background is removed from the mass spectrum and only low-IP species remain [Sch93, Kre93]. The signal-to-noise ratio dramatically improves although the ionization efficiency in the Penning regime is much lower than at the usual electron energy of about 70eV. This "soft ionization" of $Na_k$ (IP about 4 to 5eV) still has a lot of energy left over to fragment and eject the sodium cluster from the droplet and/or evaporate the droplet entirely.

The shape and thresholds of the ionization yield as a function of electron energy

(ionization yield curve) provide evidence on whether an impurity is located on the surface of the droplet or in its interior at the time of ionization. An ionization yield curve was first taken on $(H_2)_k$ and $(D_2)_k$ clusters agglomerated on helium droplets [Hen96]. The predominant Penning ionization of only the particle lighter than $^4He$ (that is the monomer of $H_2$) was then not yet attributed to a difference in location.

  $He^+$ migrates quickly $(10^{-10}s)$ to the center of the He droplet [Sch93] while the metastable $He^*$ is surrounded by a bubble and preferentially goes to the droplet's surface. The Penning channel is therefore more effective for impurities on the helium surface. As the electron ionizer energy is increased, there is a competition between the Penning $(X + He^* \rightarrow X^+ + He + e^-)$ and the charge-exchange $(X + He^+ \rightarrow X^+ + He)$ ionization channels.

# 2    Experimental Setup and Procedures

## 2.1.  Vacuum System

The set-up consists of three main chambers: Source chamber, pick-up chamber and detection chamber. The first two have each a separate roughing access so they can be worked on without cooling the diffusion pumps or having to let other chambers be contaminated by the atmosphere. Chamber pressures are monitored with Bayard-Alpert type ionization gauge tubes having single thoria coated iridium filaments. With these, air particles are 5.55 times more often ionized than helium atoms. This needs to be taken into account whenever a pressure rise due to helium is read from the gauge controllers (Granville-Phillips, Series260 and 270).

The **source chamber** is pumped with an un-baffled diffusion pump (CVC, PMCS−10C, 4.5kW power) that can be rapidly (emergency) cooled. It uses medium grade diffusion pump oil (Lesker, DIFFOIL 40) and has a pumping speed of $S_{He} = 5.6m^3/s$ for helium. A baffle would render the pumping speed too low for the high load. The diffusion pump is backed by a roots blower (Kinney, integral drive, water cooled endplates booster pump KMBD400C with 3600rpm and $S = 400cfm$) backed by a two stage rotary vane mechanical pump (Varian, SD-700).

Given the short converging nozzles for strongly supersonic beams, the Helium expansion at the source can be considered as isentropic if no diverging section is present. Then the mass flow $\rho v A$ at the nozzle exit is [Mil99]: $M = P_0 A \{[m_{He}/(k_B T_0)]\gamma[2/(\gamma+1)]^{(\gamma+1)/(\gamma-1)}\}^{1/2}$, with $r$ the nozzle radius. Applying $A = \pi r^2$, $\gamma = 5/3$, and $M = m_{He}N$, where $N$ counts atoms leaving the nozzle, one may derive $N = P_0 r^2 (9\pi/16) [5/(3m_{He}k_B T_0)]^{1/2}$. Once inside the chamber, particles equilibrate with the chamber walls and hence supply a volume $V = Nk_B T/P$ of ideal gas at $T_{r.t.}$, and this in turn is the volume being pumped, i.e. the product of pumping speed $S$ and time $t$. One derives $S = \frac{P_0}{P} r^2 \frac{9\pi}{16} \sqrt{\frac{5k_B}{3m}} \frac{T_{r.t.}}{\sqrt{T_0}}$. A typical measurement gotten with an $r = 2.5\mu m$ nozzle is $T_0 = 12K$, $P_0 = 20Bar$, $P = 5.55*2.7*10^{-5}Torr$ (for helium corrected gauge pressure), and yields 5.5m³/s. The pumps specified pumping speed is $S_{He}$ = 5.6m³/s. The same flow of gas has to be removed by the roots blower, thus $S_{diff.\ pump}P = S_{roots}P_{backing}$. The backing pressure was measured to be 5mTorr which leads to $S_{roots}$ = 0.17m³/s. Per manual, the pumping speed is indeed 400cft/min = 0.19m³/s.

The **pick-up chamber** is pumped by a water baffled diffusion pump (Varian, VHS−6) having $S_{He}$ = 3m³/s (without baffle). It must withstand any dopant we decide to pick-up with the droplets and uses silicone oil (Duniway, DS-7050-500) for chemical stability. It is backed by a mechanical pump (Welch, Duo-Seal) fitted with a copper wool filter (oil trap) against the back streaming of its oil into the diffusion pump or further. In this chamber, we want to establish partial pressures of often only few µTorr in the pick-up cells in order to pick up for example just one molecule per droplet on average. The droplets readily pick up residual gas particles encountered, too. Hence, the base pressure has to be as low as possible. After baking out the chamber walls for several days, base pressure is below $2*10^{-7}$Torr. Filling a liquid nitrogen cold trap removes $(1.7 \pm 0.3)*10^{-7}$Torr. In result, most experiments struggled at a non-helium chamber background pressure of about $(7 \pm 3)*10^{-8}$Torr with the helium beam adding $5.55*(3 \pm 2)*10^{-7}$Torr. The non-helium base pressure of roughly $10^{-5}$Pa corresponds to a particle number density $n = P/(k_BT)$ of about $n \approx 2.5*10^9$cm$^{-3}$ at $T \approx 290$K. With hot pick-up cells in the chamber, even if drawn away from the beam, the pressure will be higher. There are $L = 19.25$" $\approx 0.5$m between skimmer and entrance aperture of the alkali pick-up cell. The large He$_N$ that one needs for picking up many atoms for clusters, say $N = 15000$, will already pick-up $<k> = \sigma nLF$, that is $4*10^{-16}*3*10^{15}*0.5*2 \approx 1$ rest gas molecule on average before getting into the pick-up cell!

The **detector chamber** has as little background as possible from atmospheric gases, diffusion pump oil molecules and the doping particles introduced in the pick-up chamber. It is therefore an all conflat chamber pumped by a water cooled turbo pump (Leybold, TMP361) with $S = 0.35$m³/s for nitrogen that is backed by a mechanical pump (Welch, Duo-Seal) fitted with an oil trap as had above. After bake-out with at least 160℃ to remove water absorbed into the steel, the pressure is $P_d \leq 2*10^{-9}$Torr.

## 2.2. Source Chamber and Source

Research grade helium (99.9999%) goes into the source chamber via a 0.5micron porous sinter filter (Swagelok). It is then cooled, that is, the tubing is wrapped several times around the first stage of a cryostat's cold head (ARS, DE204SF). With a newer generation closed-cycle cryostat, our $T_0$ is usually in the range of 9.7K to 16K, thereby producing sub-critical and supercritical expansion regimes and resulting in up to $<N>$ = $5*10^5$ atoms per droplet translating into $\sigma \approx 2*10^{-15} m^2$. The closed-cycle helium cryostat's expander uses the Gifford-McMahon refrigeration cycle and is fed by a water cooled compressor (ARS-830). The cold head has two stages: the first one cools to $BP_{N2}$ and the second one to a minimum of 6K where it still has a cooling power of 3W. The first stage has no shield attached to isolate the second stage from heat radiation because at the chamber pressure of up to $10^{-4}$ Torr, the major heating is due to helium atoms bouncing between cold head and vacuum chamber walls. Instead, many layers of Al coated 6.4μm thick polyester film (Lake Shore, NRC-2 multi layer insulation) are wrapped around the completely assembled head (both stages). A solder iron melts a small hole through the layers in front of the nozzle so that the helium beam can pass through.

The second stage supports the (5 ± 1)μm diameter platinum pinhole nozzle (EMS, 42005-PT). The copper nozzle mount that is attached to the cold head is the standard "Goettingen" design with few modifications. A small Au washer, annealed for 3min at 1kK (dim red glow), is put as the gasket between a Cu knife edge and the pinhole nozzle.

Two Si diode $T$-sensors (Lake Shore, DT-470-DI-13) monitor the nozzle. One is used to PID-control the counter-heating with a 50Ω resistive heater. Counter heating may stabilize or plainly raise the stagnation temperature to yield smaller droplets. The $T$-controller (Sci. Instr., 9600-1) is either manually controlled or via an RS232C interface. The whole cold head is mounted on a $xyz\varphi$-ro−translational stage ($x$ being the setup's optical axis) in order to align the nozzle in front of a skimmer and to ensure that the beam of helium is going through all chambers and into the small entrance of the mass spectrometer. A small window in the chamber wall allows an approximate alignment of all chambers and their apertures, pick-up cells and such with a laser pointer. The solid angle of the cold cluster beam is narrower than the atomic beam. If one has not followed the cold heads contraction by adjusting the source's $z$-coordinate, it will necessitate monitoring the helium dimer signal via a phase sensitive amplifier ("lock-in"), else one may not find it. The next chamber is reached after a 0.2mm radius molecular beam skimmer (Beam Dynamics). Hence, to optimize the beam alignment, a walk-in procedure of the alignment in $y \times \varphi$ space is needed, lest one wants to be stuck at constrained extrema.

The skimmer must pierce the Mach shock disk to let only the central part of the zone of silence inside the supersonic expansion through. The distance $x$ of the nozzle to the Mach disk shock is described [Mil99] by $x^2 = (1.34r)^2 P_0/P$. Applying

$$S = \frac{P_0}{P} r^2 \frac{9\pi}{16} \sqrt{\frac{5k_B}{3m}} \frac{T_{r.t.}}{\sqrt{T_0}}$$ from above onto the results in $x = \frac{4}{3} 1.34 \sqrt{\frac{S}{\pi T_{r.t.}} \sqrt{\frac{3m}{5k_B}}} T_0^{1/4}$.

The power of ¼ makes this quite weakly dependent on $T_0$. In practice one is bound between a minimum of a few K and a high end around 20K above which the clustering of helium quickly disappears. Therefore, the distance between nozzle and skimmer is basically dependent on the diffusion pumps pumping speed alone. Since $S = 5.6$m$^3$/s, working between 6 to 20K means $x = (3.3 \pm 0.5)$cm, therefore the nozzle does not need to be adjusted for temperature if it is once aligned at about 2.5cm close to the skimmer. Hence, after the skimmer, the beam diverges by an angle of arctan(0.2mm/2.5cm) from the mid axis. After about 50cm at the entrance aperture of the first pick-up cell, the beam may diverge as far as 50cm * 0.2mm/2.5cm = 4mm from the optical axis and thus hit the inside of the scattering cell.

### 2.3. Pick-Up Chamber and Cells

The pick-up chamber is about 40cm long and ends in the automatically filled liquid nitrogen (LN2) cold trap. 10'' after the skimmer, the beam is chopped by an optical 50% chopper wheel. It is driven by a brushless DC motor (Faulhaber, 353K024BRE45) because even noble metal brushes depend on atmospheric water concentrations for lubrication. Any signal that comes with the chopping frequency (97Hz) into the detector is beam related (beam-carried if the phase is correct) and can thus be distinguished from the mostly higher but constant background signals. To yield frequency and phase, the chopper uses a slotted optical switch (Optek, OPB804) with its LED-(+)-terminal connected to its collector and the chamber wall through $100\Omega$ and $10\Omega$ respectively. The motor's and LED's (-)-terminals are shorted to the chamber. Thus, chopping adds only three wires in the chamber, which is appreciated when working with aggressive dopants like Li. With an odd number of blades, the LED is covered when the helium beam is uncovered and vice versa. Thus, detectors are triggered on negative edge ($\urcorner$).

Two with clamp heaters resistively heated Cu pick-up (or "scattering") cells can be independently translated into the beam's path with their centers at 17'' and 20'' after the skimmer. They have thermocouple sensors to monitor and adjust the temperature with controllers (Omega, CN9000A). The cell apertures have a diameter of $d = (3/16)''$. The last one is the optical stop aperture to the diverging beam and determines its cone angle to be $2\arctan[(3/32)/(21.75)] = 0.5°$. The beam diameter is therefore effectively $\Delta y_{\text{beam}} = x/116$. The wall thickness equals the inside radius $r = 2d$ of the cylindrical cells. In first approximation, the vapor pressure inside the cells is decreasing linearly down to zero along a channel through the wall. This results in an effective scattering path of length d for any cell wall. The total scattering path through a cell is thus $d + 2r + d$, i.e. $L = 6d = (2.86 \pm 0.05)$cm with the uncertainty coming from the variation of vapor flow conditions at the apertures.

While baking the chamber, the cells are kept at least 50K hotter than the chamber walls to avoid them serving as condensation sinks.

For a typical run, a cell is loaded with freshly, in hexane cut Na stick carefully minimizing exposure to air. During bake out, the cell is once heated to 270°C for a minute to crack any oxide layer on top of the liquid Na. During a run, it is stabilized anywhere desired inside the experimental range of $(180 \pm 50)$°C, giving Na vapor of up to 300μTorr ($n = 6*10^{18}$m$^{-3}$). The beam passes through the first cell that may have vapor of $10^{-4}$Torr from heated rock salt for example. After picking up molecules there, the beam traverses the second cell which contains the alkali metal. Additionally, the whole pick-up chamber can act as a further pick-up cell when opening a double needle metering valve connected to a reservoir of for example CO or solid $H_2O$.

## 2.4. Detector Chamber and Analysis Equipment

After crossing the pick-up chamber, the beam passes through a 5mm aperture at the entrance of the detector chamber. The chamber terminates in an on-axis UTI-100C quadrupole mass spectrometer ($T_{max} = 400$°C with external cables disconnected) that can operate at up to 180°C. Hence, baking proceeds at 180°C with the spectrometer operating so that its filaments do not provide condensation sinks. In order to pass laser light on axis, the UTI was replaced by an off-axis spectrometer with 500u mass range (Balzers, QMG511).

Due to the low temperature and soft surface of He clusters, desorbing guests can have very low speeds relative to the beam [Sti95, Sti96]. Therefore, the mass analyzer's e$^-$-bombardment ionizer is a long 38'' from the last cell's center so that desorbing species have time to leave the optical axis and miss the ionizer. The quadrupole mass filter ends into a Faraday cup whose ion current may be directly measured. The secondary electrons released from the cup's surface upon the ions' impacts are extracted into a secondary electron-multiplier (channeltron) who's output current feeds into the pre amplifier that belongs to the mass spectrometer's controller.

Given high residual background gas levels and low signal intensities, the output of the spectrometer's controller is mostly analyzed by a lock-in amplifier (SRS, SR510) whose phase must be adjusted or a dual channel DSP lock-in (SRS, SR830). These extract the first harmonic from signals that come in with the chopping frequency which is supplied as the reference frequency from the optical switch inside the pick-up chamber.

Measurements are partially controlled by Lab-View programs (written by the author) using a data acquisition input/output board connected to a PC card (National Instruments, PCI-6024E), GPIB and serial bus connections. The user selects the type of experiment (e.g.: "mass scan" or "ion yield curve"), inputs whether to use only the spectrometer or also a lock-in amplifier, temperature ranges and so on. The programs adjust the source temperature and all lock-in parameters (like sensitivity) and read the signals from the spectrometer, lock-in and temperature controller, alert the user via audio if manual settings have to be made, average and adjust for certain setoffs, and, after performing some statistical data analysis routines, output for example a mass spectrum showing the logarithm of the signal normalized to the initial amplification and sensitivity into a spreadsheet.

## 2.5.  Chopping and the Lock-In Technique

The mass spectrometer is very fast. Also the lock-in input impedance is 100MΩ and gives with usual capacitances connected a time constant $\tau = 10\mu s$, or in terms of an uppermost bound on the chopping frequency: $f \leq \mathbf{100kHz}$. This is beyond most of the 1/f-noise. High frequency reduces the relative error $(\sqrt{n})/n$, say expressed in $n$ as the number of chop periods per lock-in integration time. The integration over at least $n \approx 7$ cycles that the lock-in needs could be done fast, the spectra swept fast.

The above is the ideal world, now comes the real one: The original chopper had 30 blades. The beam hits the chopper wheel 5cm off-center. Hence the width of the blades at this radius is 5cm$(\pi/30)$. The chopper is $x = 10$'' behind the nozzle. The blades are thus 2.4 times as wide as the diameter of the beam $\Delta y_{\text{beam}} = x/116$. The mass spectrometer's signal of the helium dimer fragment $^4\text{He}_2^+$ is representative for the overall neutral beam intensity. Monitoring that signal with an about $<v> = 300\text{m/s}$ beam resulted in a lock-in output voltage of $U / mV = 15.7 - \sqrt{f / 67.5 Hz}$ with good statistics ($R^2 = 99$, $|t| \gg 2$, P-value $\ll 1$) in the range of 250Hz $< f <$ 2.7kHz. The upper limit was cautiously chosen to be below 100Hz of the chopper wheel's axis rotation because the motor, although rated for 8100rpm = 135Hz, might overheat in the thermally insulating vacuum. Anyways, the result shows: The slower the better! What goes wrong?

The 50% chopper removes half of the signal strength and shapes the signal into a pattern of which the lock-in extracts only the first harmonic sine (e.g.: only about 60% of a square wave's area). The minimum width of the chopper blades equals the beam width. Assume the beam cross section is circular and of homogeneous intensity. An along $y$ moving blade that now covers an $r = 1$ beam will reveal the beam's cross section area in almost linearly rising fashion: $\int 2\sqrt{1 - y^2}\,dy = \frac{\pi}{2} + y\sqrt{1 - y^2} + \arcsin(y)$.

Thus, at minimum blade width, the beam's intensity just after the chopper is almost triangular in time and can be modeled using its first harmonic $\sin(2\pi vt/\lambda)$ with $v$ being the velocity of the droplets: $I \propto [1 + \sin(2\pi vt/\lambda)]$. However, the blades must be wider because the pattern will be smeared out. The smearing may be understood as follows:

If sampled by a window of length $d = \lambda/2$, a square wave is smeared out to a more sinusoidal saw tooth pattern, so signal is gained. If cross correlated with a longer window $d > \lambda/2$, the teeth will start to overlap and signal will be lost again from the first harmonic. Hence the chopping frequency $f = v/\lambda$ has to be limited to $f \leq \frac{1}{2} v/d$. The ionizer's length is such a sampling window limiting to $v/d \approx (100\text{m/s})/\text{cm} \approx \textbf{10kHz}$.

Traveling a path of length $L$ after the chopper, the velocity distribution $\Delta v$ will smear the signal by $d/\Delta v = L/v$. This "window" limits therefore to $f \leq \frac{1}{2} v^2/[L\Delta v]$. A supersonic expansion without clustering leads to $<v> \approx 100\Delta v$ and also limits to about 10kHz. A $N = 30000$ helium droplet going at 300m/s is expected to slow down by 6m/s when picking-up 40 lithium atoms, thus impurity pick-up can smear the signal and so can the bare RMS-momentum of droplets that have evaporated many helium atoms. Even without any pick-up, because of the condensation to clusters, the speed ratio deteriorates below 30K and large droplets from a super critical expansion's fractionation can be very slow [Buc90]. There is also smearing due to the bimodal distribution which is especially bad with critical expansions. For a pessimistic estimate, assume co-expansion behavior $mv^2 = <m><v>^2$ for $He_N$ droplets. Applying the droplet size distribution's $\Delta N \approx <N>$ gives $<v> = 2\Delta v$ and thus limits $f$ to about $(100\text{m/s})/\text{m} \approx \textbf{100Hz}$. This is quite a low limit that hardly avoids line noise.

At 240Hz the lock-in was unable to lock onto the frequency because the chopper's motor only rotates with $f/(\text{\# of blades}) = 8\text{Hz}$ (when having 30 blades) and friction renders it unstable. When down-adjusting the angular speed, friction stops the motor entirely at a rotation of about 3.5Hz. The chopper was refitted with a wheel having only 5 blades. Experiments are done at $f = \textbf{97Hz}$, a prime number with little chance of being influenced by line harmonics, where stable rotation and long motor life combine. The lock-in needs about 7 chop periods of integration time $\tau$, so it is set no smaller than 7/100Hz = 70ms, which is extremely far below its capabilities.

### 2.6. Taking of Ionization Yield Curves

The ionization yield, basically the height of a species' peak in the mass spectrum, is a function of the electrons impact energy. Varying the electrons' acceleration voltage $U_{ee}$ from about 19V upwards and plotting the yield gives the ionization yield curve. With the UTI-100 spectrometer one has to calibrate for the dependent emission current $I_{em}$. The relation is linear and stable over many months for our device $[(U_{ee}/\text{V}) = 10.413 + 1653*(I_{em}/\text{A})]$. Balzers' spectrometer keeps $I_{em}$ constant. This makes the correction unnecessary but leads to burned out filaments if one is not careful and goes too far below $U_{ee} = 19\text{eV}$.

The ionization efficiency in the Penning regime is low and measurements must be taken over long times at any given electron energy. Thus, taking a whole ionization yield curve can take up to several hours for elusive species. Unstable source conditions cannot be avoided over such long times. Therefore, the desired range of electron energy and many other parameter ranges of interest must be best probed (pseudo) randomly or - second best – multiply swept. Testing whether smaller clusters from a warmer nozzle lead to different shapes of the ionization yield curves is a good example. $U_{ee}$ is set manually with a fragile potentiometer inside of the spectrometer's controller. Addressing the range randomly would overburden the potentiometer having to withstand very many turns due to large voltage steps and also the experimenter who has to turn it to random values, draining concentration. Thus, voltage was swept for every single $T_0$, the range of which was also swept trice, because of long stabilization times for the nozzle output whenever $T_0$ was set to a lower value. In order to still test at many different points all over the ranges and to have these supports at a constant density along the ranges (homogeneous covering), I made Labview sweep a range min $\leq m \leq$ max of property $m$ as described in the following employing mnemonics:

The computer counts $n$ from 0 to ($N$-1). Variables written with capital letters are often having large values. They are "heavy and at the bottom", like 1/$N$ [German(de**N**ominator) = **N**enner] and 1/$F$ [German(**F**loor) = **F**lur, also something that is **S**wept]. 1/$N$ := 1/$S$ * 1/$F$. The "**f**(loor) size" is $f$:=(max-min)/$F$ and the **s**tep size $s$ := $f$/$S$ is $s$ = (max-min)/$N$. Any programming language provides easy access to the modulus mod[$n/F$] of a fraction $n/F$. The integer part is int[$n/F$] := ($n$ - mod[$n/F$])/$F$.

With these definitions it is easily seen that $m$ = min + $s$*int[$n/F$] + $f$*mod[$n/F$] will sweep the range $S$ times, every time visiting each floor once like on an elevator, each stop stepping out to add another measurement one step size $s$ above the previous time on that floor.

E.g.: 26.5V $\leq U_{ee} \leq$ 86.5V was swept $S$ = 12 times for every $T_0$, the range of which is 7.2K $\leq T_0 \leq$ 10.2K and is swept $S$ = 3 times. $F$ = 5 floors in both ranges led to $N$ = 60 and $N$ = 15 yielding a total of 900 data pixels ($U$, $T_0$).

## 2.7. UV Light Setup

The detector chamber was fitted with two (entrance and exit) silica (suprasil) windows with (0.86 +/- 0.01) transmittance to vacuum UV due to reflectance. They restrict bake-out to 200°C (lead-silver braze alloy MP = 300°C). A water cooled 1kW arc lamp (PTI, A6000) is mounted together with all filters and optical baffles on a platform on three *z*-elevation screws for height and tilt adjustment that sit in turn on a *xy*-translation stage. Ionization of Li and Na atoms requires wavelengths $\lambda_{IP} = 230$nm and 241nm respectively. These are shorter than the threshold of ozone production at 242nm. A non-ozone free short arc bulb (HgXe 6293) is used and the ozone from the lamp housing and attached optical baffles vented to outside of the laboratory. The lamp's $f = 4.5$ elliptical mirror collects 60% of the bulb's light in the external focal spot. Given the bulb's efficiency, without filters about $(12 \pm 2)\%$ of 38V*30A are delivered to the about $(3\text{mm})^3$ focal spot that we can shift into the modified cage of the electron bombardment ionizer of the mass spectrometer by adjusting until the cage is in the fields of view of two narrow sight tubes glued to the lamp housing so that their optical axis cross each other in the focal spot.

IR-light is removed with a flowing tap-water cooled filter (Oriel) containing de-ionized water. Commercially available distilled water contains iron ions from pipelines transferring the water after distillation. Iron ions' water complexes absorb UV extremely well. The filter is made from metal and should be passivated and fresh water added just before use. Measurements with an UV laser and a SiC UV-photodiode (JEC 0.1 S) confirmed that the water between the filter's suprasil windows reduces their reflectivity. Low power visible light was absorbed in a solar blind filter (Corning 7-54) positioned inside the water filter for cooling. It is destroyed in 10 min at the power needed to attain the bulb's specified spectrum and life time. Even less IR and more as well as even shorter wavelength UV can be had when substituting with 66mg/ml high purity $NiSO_4$ in the water filter. Spatula, funnels and scale must be non ferrous.

# 3    Experimental Results

## 3.1.  Motivation: Cold Metallic Alkali Clusters and Wetting Behavior

It is not easy to cool clusters, which mostly come from hot supersonic expansion sources, in flight. One can aggregate quite big clusters inside $He_N$, for example $Ag_{1-150}$ [Bar96, Fed99], large Indium clusters [Bar96] and huge clusters of Mg [Die01, Dop01] yet not alkali. Hot alkali clusters have been extensively studied [Deh93], especially Li and Na clusters are well investigated because they are simple one-valence electron atom systems and give insight into general cluster physics easily. $Na_k$ is so to say the hydrogen atom of cluster physics. In particular, extensive spectroscopic studies of electronic excitations in sodium clusters have been performed [Bla92, Rei95]. The origin and character of line shapes of these species are still being discussed. The long life times ($\approx$ 10 times $\Delta t = \frac{1}{2}\hbar/\Delta E$) of collective electron excitations (plasmons) in these clusters is still not understood. Spectroscopic studies of electronic transitions have been carried out at low temperatures [Hab99], where it is reasonable to expect that the physics is not obscured by averaging over thermal ensembles. The coldest reported spectra for medium-sized free sodium clusters in a molecular beam are at 35K for $Na_{11}$ [Ell95] but still appear to be noticeably broadened by thermal effects. Deposition on a cold substrate can give lower temperatures, but the spectra of matrix-deposited clusters suffer inhomogeneous broadening and some valence electrons are donated to the matrix, shifting the magic numbers.

As stated, alkali clusters have been produced on helium droplets and they can stay attached to the surface [Sch04, Von02] yet they seem to be only weakly (vdW) bound clusters (not metal clusters) because the release of the large covalent (metallic) binding energy leads to complete evaporation of the droplet or immediate desorption from the surface. It was even anticipated [Leh99, Hof99] that it is impossible to grow larger then $Na_3$ particles by the pick-up method, because the energy liberated upon condensation of $Na_4$, even if spin polarized and only vdW-bound, would be sufficient to eject the particle from its weakly bound position on the surface dimple. We found though even much larger $Na_k$ particles [Von02]. Their nature and the mechanism of their growth is still not fully understood [Von03, Bue06].

Helium does not wet cesium – not even bulk Cs. However, if Cs is deposited on gold [Tab94] up to a layer thickness of 20 Cs atoms, $^4$He will wet the Cs surface. Such studies [Tab94, Her97] are very sensitive to surface contaminations, thus also showing that wetting behavior depends much on seeds even if they are far away from the surface in question. This induced wetting cannot be an attraction all the way through 20 layers of cesium. Work function and ionization potential of Au are much larger than those of alkali. The gold takes up valence electrons from the cesium and thereby reduces the heliophobe electron spill-out beyond the surface. This means that the seed must contact the alkali *first*; only then is wetting thereby induced. Attraction through helium *before* any contact may be existent for xenon cluster seeds of size $k > 14$ dragging slightly heliophobe barium atoms from the surface [Lug00], but even those results can also be explained by the barium penetrating quite close to the large xenon cluster upon impact.

Nevertheless, large attractive vdW interaction between alkali and some potential chromophores had been reported [Kre93, Kre98]. Therefore, we tried to pre-seed $^4$He clusters with attractive, wetted atoms and molecules like H2O, NaCl, HI or CO such that metallic alkali clusters may grow where the seed resides inside the helium.

Helium wets bulk sodium but the ion yield curves prove that the sodium fragments detected originate on the droplet's surface. If metallic $Na_k$ clusters are not wetted and stay on the surface, there must be a novel wetting transition depending on $k$. Such has not been suggested by any theoretical approach except calculations indicating that alkali monomers and dimers stay on the surface [Anc95, Ler95]. Moreover, the heliophobe electron spill out beyond the surface of bulk sodium is no less than the spill out for metallic $Na_k$ with $k$ being as large as for a large fraction of the detected sodium fragments originating from the helium droplet surface. Of course, the polarizability of the clusters increases with their size but also small clusters have been shown to have unexpectedly large polarizability.

(A straightforward way to investigate the properties of clusters is by measuring their electric polarizability [Bon97]. A perfectly conducting sphere of radius $R$ has polarizability $\alpha = [4\pi\varepsilon_0]R^3$ proportional to its volume. The polarizability per atom is $\alpha_{per\_atom} := \alpha/N$. The Wigner-Seitz radius is defined via the bulk limit $r_S^3 := R^3/N$, which yields: $\alpha_{per\_atom} = [4\pi\varepsilon_0]r_S^3$. Large clusters do approach this limit but metal clusters show a greater polarizability for small sizes [Kni85, Tik01]. The discrepancy is explained by the electrostatic screening beyond the classical boundary [Lan70] due to e⁻-spill-out on the order of $\delta = 1.3$Å for sodium. It holds therefore $\alpha = [4\pi\varepsilon_0](R+\delta)^3$ and hence: $\alpha_{per\_atom} = [4\pi\varepsilon_0](r_S+N^{1/3}\delta)^3$. For $Na_{20}$ the correction to $\alpha$ is 40%. There are further polarizability enhancing effects though. Mass abundance spectra of $Na_n$ [Kni84] evidence a shell structure in the electronic levels that causes closed shell ("magic") sizes to be more stable than others. Closed shells occur at $n = 8, 20, 40, 58 \ldots$ . The spherical jellium model predicts larger $\alpha$ of open shell clusters compared to closed shell ones [Eck84] and even underestimates those by 15%. Corrections to $\alpha$ have been reproduced closely for clusters up to $Na_9$ with spin-dependent local-density approximation [Mou90].)

### 3.2. Mass Abundance Spectra

The spectra display a noticeable odd-even oscillation in the strengths of metal cluster peaks. Such abundance oscillations are well known in the mass spectra of free simple-metal clusters where they are understood to reflect the higher stability of even-electron systems in the electronic shell model [Deh93, Deh87].

A very strong peak at $Na_9^+$ is due to the strong binding of the closed-shell 8 electron clusters and the corresponding weakness of the $Na_{10}^+$ that likely decay to $Na_9^+$. The odd-even oscillations of peak strengths are extremely stable. They can be revealed even at low cell temperatures (e.g. $T_{Na} = 155$°C), where the sodium cluster mass peaks are buried in the background noise. For that, we look at the ratios of mass signals [Von02].

A similar odd-even pattern has been seen with silver clusters in helium droplets

[Bar96, Fed99] and cited as evidence for considerable fragmentation of picked-up clusters accompanying ionization by 70eV electrons. Our observations demonstrate that Penning ionization detection produces the same feature with sodium clusters at the surface of helium droplets. At our high doping levels the monomer to dimer signal ratio turns out to be cell temperature independent at $U_{\mathrm{Na2}} = 0.67*U_{\mathrm{Na}}$; $U$ being the lock-in's output voltage. This evidences that fractionation upon ionization is more important than any odd even effect due to evaporation processes during pick-up since odd/even magic clusters fragment predominantly by expelling monomers and dimers [Bré88, Bew94, Kru96, Heř97]. In any event, the very existence of an odd-even oscillation in metal cluster intensities implies that they arise from more complicated dynamics than just a sequential pick-up of atoms. In the latter case, the intensity ratios would be dominated by Poisson probability. Such a function would produce a monotonous variation of metal cluster signal with size $N$.

The measured intensity ratios between successive cluster sizes do not follow the Poisson distribution. For instance, if we assume Poisson behavior, the trimer intensity $I_{k=3}$ and pentamer intensity $I_5$ are correlated via $20I_5 = I_3(\sigma nLF)^2$, and the vapor density $n$ makes this exponentially dependent on the cell temperature ($T$-dependence of $F$ is negligible). The experimental value of the intensity ratio is constant at $I_3/I_5 = 2.1 \pm 0.5$ for all temperatures above $165\,°C$ (like seen for the monomer/dimer ratio). The constant ratios of mass peaks does *not* follow from magic stabilities that determine fractionation patterns at detection but instead follows from spin desorption statistics, as will be shown in the section "Theoretical III".

It is interesting that odd-numbered $\mathrm{Na}_{3 \leq k \leq 9}^{+}$ cluster ions (even number of electrons) have abundances of the same order of magnitude. Indeed, the pentamer is easier to detect than the tetramer. That the presence of the tetramer had not been recognized previously is probably due to its lying at the same mass spectrum position as $\mathrm{He}_{23}$ and its being strongly suppressed by the odd-even effect. When gradually increasing the pick-up cell temperature while monitoring the first four sodium cluster sizes by high-energy electron ionization or by laser fluorescence rather than the Penning technique, a detectable tetramer signal would not be seen until the cell temperature far exceeded the value expected for statistical pick-up of four metal atoms. This probably led to people assuming that pure helium droplets cannot support sodium clusters larger than the trimer. In fact though, at cell temperatures above 190 ℃, the intensity of $Na_{13}^+$ outweighs that of $Na_{11}^+$ and leads to conclude that these fragments originate from even much bigger clusters.

The mass spectra also reveal the presence of fragment complexes $Na_kHe_N^+$, $Na_kH_2O^+$, $Na_kHO^+$ and $Na_kO^+$. The last three are associated with the background water vapor. These peaks are very weak.

### 3.3. Guest Cluster Location

Both, thresholds and shapes, especially the maxima around $U_{ee} \approx 30eV$, of the Na and $Na_2$ ionization yield curves show that Penning ionization contributes significantly to the production of the sodium cluster ions. This is characteristic of a surface location.

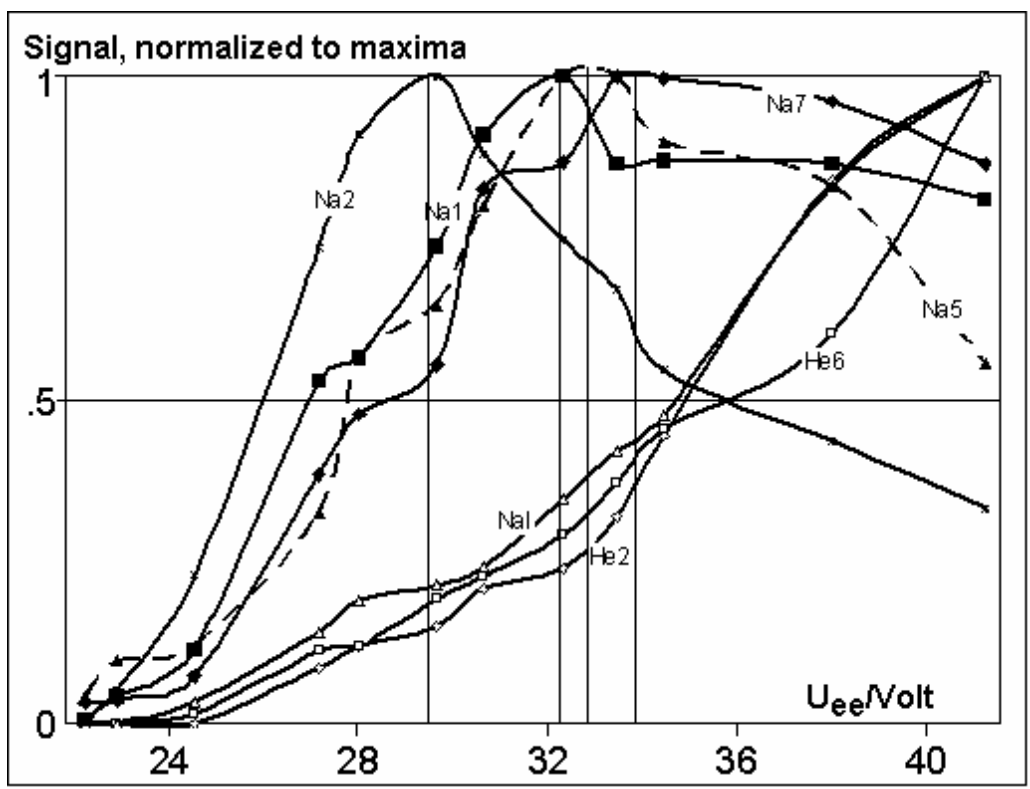

Ion yield curves of dopant ions and He$_2^+$ and He$_6^+$ fragments [Von02]: The electron energy scale (given as acceleration voltage $U_{ee}$) has been shifted by setting the He$_2^+$ threshold to the free atom's IP at 24.6eV. The zero point of the instruments voltage was not well enough calibrated to discern the energy barrier encountered when electrons enter the helium droplet.

The thresholds and the monotonic growing of the ion yield curves of He$_2$, He$_6$ and of the solvated polar molecule NaI are quite different and reflect the position inside the droplet.

The Na$^+$ and Na$_2^+$ signals are always much stronger than all other sodium related signals. They must derive from and are therefore representative of larger Na clusters having fractionated. One may conclude that all of the alkali clusters, even including much larger ones than the spectrometers mass range allows to detect, have been located and ionized at the droplet surface. We could observe this directly even for the penta- and heptamer ion yield curves (see figure).

### 3.4.   Nature of the Guest Clusters' Bond

#### 3.4.1.   Fragmentation

Our investigation showed that alkali clusters grown on helium fragment substantially even under soft Penning ionization. This partly motivated our trials at even softer UV photo ionization. Moreover, a large proportion of the fragments are sodium dimers. Hot metallic $Na_k$ are known to evaporate also dimers [Bré88, Bew94, Kru96, Heí97] because of the odd/even alternation (along $k$) in binding strength due to pairing of shell electrons. The huge dimer fraction observed though is more reminiscent of either an enrichment of spin polarized small vdW bound clusters due to desorption statistics [Von03] and/or a somewhat violent chain reaction after an initial spin flip in a larger vdW bound cluster.

#### 3.4.2.   UV-Ionization

The UV experiments never led to any signal. Given the sizes of the fragments as monitored by the mass spectrometer when in electron bombardment mode, we should have seen ready ionization via UV if any metallic sodium clusters had been present. Given that enough short wave length UV was present to ionize even atoms, the absence of a monomer signal is suspicious. The UTI mass spectrometer is "space charge operated": The charge density gradient of the cloud of electrons pulls ions onto the optical axis and also accelerates them to the aperture electrode before the actual extraction voltage behind that aperture can grab them. It is likely that with cold filaments and UV light instead of electrons, not enough ions can be focused into the quadrupole filter's aperture.

### 3.4.3.  Evaporation versus Desorption

The large alkali-alkali metallic binding energy is enough to evaporate the whole helium droplet. Any energy $E$ released *inside* a droplet equilibrates very fast over the whole super fluid cluster so that evaporating surface atoms take little more away than the small He-He$_N$ binding energy $e$; the number of evaporated atoms is thus always close to the maximum $E/e$. However, the surface location of the clusters promised that atoms evaporated shortly after the release of energy originate from close to the sodium in a region that is turned normal fluid and already at the surface. If a substantial part evaporates locally and before the energy $E$ had time to equilibrate over the whole droplet, than the evaporated atoms take more energy away than $e$. Helium might expand explosively in between the sodium and the rest of the still cold and super fluid droplet, thereby separating them. This desorption channel has been observe

### 3.4.4.  Seeding Experiments

Spectroscopy will be complicated by a seed's presence but using xenon as a seed to enhance the pick up of barium [Lug00], spectra shifted very little. A wetted impurity carries a dense helium solvation shell that keeps heliophobe species away. A seed's influence, even if close to an alkali cluster, can also often be accounted for. The physics of small alkali clusters is determined by their electronic shell structure and studies of cluster impurities [Mal89] showed that the seed molecule just takes up a few valence electrons (e.g.: oxygen takes two) and therefore just shifts the properties of the Na$_k$ by as many steps along the $k$-axis.

There is involuntary seeding from background rest gasses. Sodium attached to water seemed to appear but could not be positively identified as sodium being drawn to a water molecule inside the droplet. Fractionation of clusters can be quenched by shells of added molecules like CO. Trying to quench the strong fractionation of the sodium clusters in such a way did not result in any mixed $(CO)_j Na_k$ fragments. $(CO)_j$ fragment peaks could be seen interspersed with peaks of $Na_k$ fragments.

Trying to detect just sodium monomers fixed to the seed with the largest electric dipole moment, namely CsCl, has been fruitless. In earlier experimentation with NaI without addition of sodium, the pure salt seeds showed up as $He_{0 \leq n \leq 3}Na^+$, $(NaI)^+$, $He_{0 \leq n < 18}I^+$ and $(Na_2I)^+$. A $(Na_3I)^+$ fragment could not be detected. Since NaI is a salt and strongly ionic ($Na^+I^-$), this absence is expected due to a predominant $(Na_3I_2)^+$ channel from trimers. Whether there were not enough trimers to start with could not be established because $(Na_3I_2)^+$ is beyond the spectrometers 300u mass range.

After adjusting the salt temperature to maximize for the pick-up of on average only one NaI seed, the further pick up of sodium by seeded helium clusters resulted in $(Na_{1 \leq k \leq 6}I)^+$. The signals of $(Na_kI)^+$ did have maxima at vapor pressures below the maxima of the corresponding $Na_{k-1}^+$ fragments of experiments without pre-seeding, therefore pickup was indeed enhanced or desorption reduced. The odd/even alternation proceeds qualitatively as expected, i.e. as if the seed is not present. We could not establish whether $(Na_kI)^+$ was originating from the interior or the surface of the droplet. The only data could be taken at low ionization energies were the surface related signals are maximal.

One may also ponder the strength with which the iodine ion holds on to many helium atoms while almost no helium stays with sodium containing ionic species. Ion yield curves show that the salt seed clusters are ionized inside of the droplet. If $I^+$ results from ionization of NaI, the heavy iodine ion will stay in the center while the sodium rest might be getting enough momentum to leave the droplet. Why do the even heavier $(NaI)^+$ and $(Na_2I)^+$ not hold on to helium atoms?

There was no thorough enough exploration of the parameter space to pin down whether for example seeding with CsCl ever led to any mixed species having both sodium and chloride or both sodium and cesium. Changing the stagnation conditions, e.g. heating the nozzle, leads to smaller droplets. At some point it is inconceivable that the seed in the center was not close enough to the surface in small droplets to be able to attract a further Na atom impinging. The strong desire of cesium to get rid of its valence electron may have led to any mixed species always separating the added sodium upon electron impact. The fractionation of $(NaI)_k$ upon ionization might be a misleading guide to predicting the fragments of electron impact ionization of $Na(CsCl)_k$. Moreover, the ionization probability is proportional to the droplets cross section and conspiring with the small systems more likely deflection from the beam path against signals due to small droplets.

### 3.5.   Summary, Outlook, Suggestions

Producing ultra cold metallically bound sodium clusters was maybe unsuccessful except for the seeding experiments showing $(Na_{1 \leq k \leq 6}I)^+$. A detectable fraction of metallic alkali clusters on helium droplets has never been clearly identified. If there are such clusters on or in a sizable fraction of droplets, they will lead to very strong beam depletion if illuminated with visible light because of the very high absorption cross section of the plasmon resonance of alkali clusters. This beam depletion experiment is now set up with an argon ion laser. Its output would not affect the alkali atoms, neither at the ionization energy nor at the respective D-lines that have practically all the oscillator strength, but it would pump up almost all metallic alkali clusters plasmon resonances. Nevertheless, depletion signals could be argued as due to the weakly bound spin polarized clusters, the properties of which are unknown. Also, the presence of a seed complicates things, hence laser probing a reestablished $(Na_{1 \leq k \leq 6}I)^+$ signal is not a short term project.

The following are suggestions of experimental routes and theoretical attempts to deduce the right parameters needed for producing covalent alkali clusters on helium:

### 3.5.1.   Aggressive Doping

I define "aggressive doping" via $\tau < T_{surface}$, where the mean collision free time $\tau =$

$L/(<k> <v>) = 1/(nF\sigma<v>)$. It means that along a very short pick-up path $L$ with high vapor density $n$, heliophobe atoms impinge so rapidly that the mean impact free time $\tau$ is shorter than what it takes developed bubblons to *b*reach through the exit barrier at the droplet's surface. Alkali bubblons would than meet inside large helium droplets and metallically binding alkali would not be expelled from the surface but likely stay inside.

Let us first argue against this: Calculated from average beam velocity $<v>$ and the strongest scattering gas density $n$ that we ever reached, we never went below $\tau \geq 10^{-7}$s. The transit time for a bubblon through a droplet is on the order of droplet diameter over Landau velocity, which puts the time to reach (not *b*reach at) the surface at only about $10^{-10}$s. One cannot get $\tau = L/(<v> <k>)$ as short as this, because $<v>$ is more or less constant, much larger $<k>$ will destroy all droplets, $L$ is already only a few cm, $L$ below the size of the scattering cell's apertures cannot support a defined vapor pressure anymore and an $L$ of about a few μm basically describes an experiment shooting atom clouds or already clusters into the helium droplets.

On the other hand, electron bubblons are also only metastable but will travel forth and back through the super fluid droplet many times before finally bursting at the surface. Electron bubblon life times of about 60ms have been measured with 22eV electrons entering large droplets $^4\text{He}_N$ with $N = 10^6$ [Far98, Kha89]). Sodium atom bubblons could last longer if the picture of tunneling out through the surface barrier is correct, because firstly, the mass $m_{Na}$ is about $4*10^4$ times that of an electron and secondly, the barrier height should be also larger. The shorter radius of a Na-bubblon, its smaller effective mass, different surface energy and so on make an estimate though non-trivial. How efficiently impinging Na atoms often even lead to bubblons inside $\text{He}_N$ is not known and must be estimated from observations with bulk He.

### 3.5.2. Ionic Expansion and Ionic Seeds

Once an ionic seed is embedded, it will increase the embedding cross section of the whole droplet. The droplet may readily pick up more atoms to be dragged into its center by the ion-snowball. Spectra of for example the $Na_k$ clusters would be shifted by the missing/addition of a few valence electrons. Tiny amounts of such $He_NA_k^{+/-m}$ will be detectable because the detectors ionizer may be turned off resulting in no disturbing background signal, and the ionization cross section that decreases with decreasing droplet size is now effectively replaced by 100% ionization probability.

Ionic expansion should be possible with a β-decaying foil surrounding the stagnation region or placed inside of it, because many decays lead to thousands of helium ions. Some ions should make it into the expansion region resulting in condensation of huge helium droplet ions. These droplets carry $He_2^+$−snowballs that will ionize the first picked up guest atom long before the usual average impact free time τ between successive pick up events. The energy release associated is almost as large as $IP_{He}$−$IP_{guest}$, but the droplets may be extremely large now and able to evaporate this energy without the acquired rms-momentum deflecting them from the beam path.

Ionic seeds may be produced by Langmuir-Taylor ionization of alkali atoms at a hot platinum helix, i.e. a Pt wire curled around the beam path inside of a pick-up cell. Now there is no large energy release associated. If sufficient flux of ions necessitates such high temperature and visible surface area of the wire that the vapor's MB distribution is altered unacceptably, a pre-seeding cell may be needed. If the helix is not too densely wound, high positive potential on its wire will increase the ion flux but the stray fields worsen the deviation of ions from the optical axis. Therefore, potential and heating needs to be periodical in time or even pulsed and could replace optical chopping.

Ionic seeding might also be possible with a β-decaying foil surrounding the entrance of the scattering cell. Droplets hit by electrons are destroyed, but the cloud of helium fragments might introduce many ionic fragments into the pick up path, and they can ionize single metal atoms and other droplets. The discussed energy release is now a mayor problem though.

# 4    Theoretical I: Strong Depletion

## 4.1.  Self-destructing Beams

Increasing the pick-up cell's vapor pressure, one encounters unexpectedly soon the limit at which the signal due to helium clusters disappears. Entering average droplet sizes into equations for the number of picked-up particles, amount of evaporated helium and so on, results in a much higher limit on the vapor pressure. Is the effect entirely due to deflection being neglected?

It was suggested [Wit00] that a narrow tube ("Wittig tube") around the beam might enhance otherwise not detectable beam depletion. A few energy dissipating events, maybe triggered by an interaction of a laser beam with guest impurities in the He droplets, would lead to evaporated helium atoms that travel to the tube's walls close by, where they are thermalized to the temperature of the tube. After returning fast to the beam, they would lead to collisions that evaporate much more helium than the triggering interaction did. The atoms from the secondary evaporation would also be heated at the tube wall. In a sufficiently narrow and long tube, a runaway process develops. This should lead to a steady state of a beam of helium clusters being destroyed inside the cloud of atoms that the beam thereby continuously replenishes. Could something similar be happening inside pick-up cells and explain the sudden loss of the signal?

The following model is applicable to many of the usually employed pick-up cell designs all the way to the Wittig-tube. This calculation is programmed into a Mathematica5® [Wolfram Research] notebook.

### 4.2. Model Geometry

A pick-up cell's interior volume consists of three cylinders attached to each other in sequence: The entrance cylinder, the main cell cylinder and the exit cylinder, which has the same proportions as the entrance cylinder. Assume that the interior of the main cylinder of the pick-up cell has a well defined pressure $P_2$, i.e. it contains a Maxwell-Boltzmann distributed vapor leading to the same pressure reading whichever direction a pressure gauge might be pointed at. The main cylinder in the middle has a radius $r$ and a height $h$. The height $h$ has to be measured from the dopant fill level upwards, because we are interested in the volume of vapor only. The other two cylinders have a diameter $d$ and length $l$. Thus the entrance and exit apertures have an area of $A = \pi(d/2)^2$. Assume that the flow conditions are molecular and that the pressure $P_2$ goes linearly down along the outer cylinders to meet the vacuum chamber's pressure $P_1$ at the entrance and exit apertures.

The scattering cell model's domain of applicability inside the parameter space $d \times h \times l \times r$ is not affected by the orientation of the main cell body. If the cylinder is put on its side, the height $h$ will add to the lengths of the other two cylinders. If we now shrink the radius to $r = d/2$, the pick-up cell turned into a Wittig tube of length $Z := 2l+h$. This provides a continuous range of models $(h, l, r)$ from $(Z, 0, r)$ to $(0, Z/2, r)$, that all describe the Wittig tube. The models differ in that the central part of length $h$ is the one with a well defined pressure $P_2$. This is the interaction region, because the initial interaction will evaporate He atoms all along it and thus lead fast to what can be taken as the initial pressure $P_2$. Any additional pressure is due to the secondary evaporation induced by the presence of $P_2$. For example: The originally proposed 100mm long tube has a radius of 2mm. The usual laser focal area of 1mm$^2$ and an opening angle of at most $\arctan(r/Z)$ make for a long interaction region $h$.

### 4.3. Particle Losses

The entrance and exit apertures have an area of $A = \pi(d/2)^2$. The transport through a surface $A$ is generally written $\mathrm{d}(PV)/\mathrm{d}t = C(P_2 - P_1)$ with $C$ being a constant dependent on the shape of the area. Let us concentrate on the practical: $P_2 \gg P_1$. $P_2 = n_2 k_B T$ with $n_2 = N_2/V$ holds only inside the main cell body $V = \pi r^2 h$. The loss of particles out of a single aperture (e.g. the entrance to the cell) is therefore:

$$\dot{N}_2^- = \frac{-(P_2 V)^{\cdot}}{k_B T} = \frac{-C P_2}{k_B T}$$

The factor $C$ is $C = \frac{A}{4}\bar{v} a_{(l/d)}$ ; $\bar{v} = \sqrt{\frac{8}{\pi}\frac{k_B T}{m}}$ and $a_{(l/d)}$ is the probability of transmission through a short cylindrical tube which cannot be calculated analytically. There are approximations but none convenient for the range that is of interest later on, especially, $(l/d) = 0$ needs to be included without leading to division by zero problems. We desire a formula rather than table values also for the computer program. A somewhat physically justified ansatz is:

$$a_{(x)} = \left(1 - \left(\frac{x}{1+x}\right)^a\right)(1 + bx - cx^2)$$

Inside the range $0 \le (l/d) \le 5$ it deviates at most by 0.2% from tabulated [Oha89] values if $(a, b, c) = (96.35, 6.04, 0.47)/100$. Including both apertures ($A_{\text{tot}} = 2A$) the formula for particle loss becomes therefore the following closed expression:

$$\dot{N}_2^- = -a_{(l/d)}\frac{A_{\text{tot}}\bar{v}}{4}n_2$$

### 4.3.1. Carry out of Dopant is negligible

How many particles $N^*$ of the scattering dopant vapor are taken out of the cell by the clusters picking them up? An aperture of area $A$ being a distance $R$ away from the source nozzle subtends a solid angle $\Omega = A/R^2$. The flux of condensed helium atoms per solid angle is $I$, and the flux of helium clusters approximately scales as $I/<N>$. Even assuming very high flux ($I = 10^{20}$Hz/sr), the speed $-\left(\dfrac{\dot{N}_2^*}{A_{\text{tot}} n_2}\right) = \dfrac{I}{R^2 \langle N \rangle^{1/3}} F \pi r_s^2 L$ is of the order of 1m/s and thousands of times smaller than the speed of loss due to particles leaving the cell *unassisted*, $\left| \dot{N}_2^- / \left( A_{\text{tot}} n_2 \right) \right| = a_{(l/d)} \left( \bar{v}/4 \right)$.

### 4.3.2. Corrected Vapor Pressure (unassisted loss of dopant)

How much does the constant loss of dopant particles decrease the pressure $P_2$ from the saturated equilibrium vapor pressure? The loss equations can model the supply $N_2^+$ of dopant particles from the liquid dopant at the bottom of the cell. The liquid has a surface of $\pi r^2$ and is "connected" via a cylinder of zero length, i.e. $a_{(l/d)} = 1$. The dopant surface supplies atoms at the listed saturated vapor pressure $P_{\text{sat}}$. The loss $N_2^-$ is due to particles hitting the same dopant surface plus the loss through the two apertures. This gives a differential equation for the number of dopant particles $N_2$ inside the cell:

$$\dot{N}_2 = \frac{\bar{v}}{4}\left( \pi r^2 \left( \frac{P_{\text{sat}}}{k_B T} - n_2 \right) - a_{(l/d)} A_{\text{tot}} n_2 \right)$$

At large time $t$, no other surfaces have net flow, the cell is at a steady state, and the overall particle number constant. That leads to a correction of the vapor pressure inside the cell:

$$n_2 = \frac{\pi r^2}{a_{(l/d)} A_{\text{tot}} + \pi r^2} \frac{P_{\text{sat}}}{k_B T}$$

Whether negligible or not, this might as well be taken into account. The improved loss formula for the dopant particles is:

$$\dot{N}_2^- = -\frac{a_{(l/d)} A_{\text{tot}} * \pi r^2}{a_{(l/d)} A_{\text{tot}} + \pi r^2} \frac{P_{\text{sat}}}{\sqrt{2\pi m k_B T}}$$

The saturated vapor pressure $P_{sat}$ depends on the temperature via the Antoine equation. For sodium the following supposedly holds with an accuracy of 5% or better all the way from the melting point up to 700K [Wea04]: $P_{Na} = 10^{4.71-5377\,\mathrm{K}/T}\ Bar$

## 4.4. Mean Free Path

The mentioned Mathematica5 notebook checks automatically whether the assumption of molecular flow is valid by calculating the mean free path $\lambda$. This is often forgotten, and using the program to estimate the expected pick-up in another experiment using a small water pick-up cell actually revealed that the researcher had left the molecular flow regime, thus the pressure in the cell was lower than calculated from its temperature.

Consider the Poisson distribution of the pick-up process $<k> = \sigma n L F$ or the exponential distribution of accident free times. For the mean free path we put $L = \lambda$ at an average of $<k> = 1$ collisions. Thus, $\lambda n F \sigma = 1$ with $n = N/V$ again the particle number density and $\sigma$ the cross section as shall be approximated with the particles' radii. For atoms, use the atomic radii. The mean free path of a gaseous *species* with mass $m_{Sp}$ inside another gas of particles with mass $m_B$ is dependent on $F^2 = 1 + m_{Sp}/m_B$. The mean free path formula for a single species is therefore easily remembered as having $F = \sqrt{2}$. Anyways, one derives:

$$\lambda_{Sp \to B}^{-1} = 4\sqrt{2}\pi r_{Sp}^2 n_{Sp} + \sqrt{1 + \frac{m_{Sp}}{m_B}}\,\pi\left(r_{Sp} + r_B\right)^2 n_B$$

At $T_{Na} = 493$K, $r_{Na} = 3$Å with the corrected vapor pressure in our pick-up cell yields $\lambda_{Na \to Na} = 3.48$cm. This is larger than the longest path inside the cell and so the assumption of molecular flow is justified. The mean free path of sodium inside helium can be smaller than 3cm if the helium has densities above $1.4*10^{19}$m$^{-3}$ (such densities will show up later on). Still, the flow is molecular because the helium gives the sodium no velocity bias. On the contrary, because at that point, $\lambda_{He \to He} = 16.9$cm.

### 4.5. Evacuation Time

One might be interested in whether the time of one chopping period is enough to empty the scattering cell of all cluster particles that the beam introduced. The time it takes to evacuate the Wittig tube is of critical importance. If it is too small, atoms will not stay around long enough to make a runaway process happen.

The probability of encountering an exit is equal to the area $A_{tot}$ of the exits over the area $A_{in}$ visible inside, which is $2\pi r(r + h)$ for a cylinder. To get the total probability of leaving the cell, this is multiplied by $a_{(l/d)}$: $\text{Prob} = a_{(l/d)}\left(A_{tot}/A_{in}\right)$. Having a try at exiting every $\delta t = L/\overline{v}$, the average time to leave is $\tau = L/(\text{Prob}\,\overline{v})$ (neglecting the time of being inside the exit, which makes it a rough approach for the Wittig tube):

$$\tau = \frac{A_{in}L}{a_{(l/d)}A_{tot}\overline{v}}$$

The time is proportional to a characteristic volume $A_{in}L$. Putting the length $L = 2rh/(r+h)$ assures $L = r$ when $h = r$ and also when $h \gg r$. The inner volume and area of the cylinder obey $V = \pi r^2 h$ and $A_{in} = 2\pi r(r+h)$, thus:

$$\tau = \frac{4V}{a_{(l/d)}A_{tot}\overline{v}}$$

This formula can also be derived via the equations for the pumping speed. $\dot{N}^- = -a_{(l/d)}A_{tot}\frac{\overline{v}}{4}n$ with $n = N/V$ and $\tau$ from above leads to $dN^-/N = -dt/\tau$, which must be integrated with $\int_0^t dt = \int_{N_{(0)}}^{N_{(t)}} dN$ to gain $N_{(t)} = N_{(0)}e^{-t/\tau}$. Defining $\tau_{P_2/P_1} := \tau \ln(P_2/P_1)$ shows that the $\tau$ derived is $\tau_{1/e}$. The example from above, i.e. $T_{Na} = 493$K in our pick-up cell, leads to evacuation times of 4.23ms for sodium and 1.77ms for helium.

### 4.6.  Gains

An aperture of area $A$ being a distance $R$ away from the source nozzle subtends a solid angle $\Omega = A/R^2$. The flux of condensed helium atoms per solid angle is $I$, and the flux of helium clusters approximately scales as $I/<N>$. If the whole beam disappears inside the cell, the gain of atoms will be $\dot{N}_2^+ = I\Omega$. If the beam is only partially depleted, the size distribution $(\mathrm{d}C/\mathrm{d}N)_{(N)}$ with $\langle N \rangle = \int_0^\infty N(\partial_N C)\,\mathrm{d}N$ must be taken into account:

$$\dot{N}_2^+ = I\Omega \int_0^\infty (N - N_{(L)})(\partial_N C)\,\mathrm{d}N / \langle N \rangle$$

$N = N_{(0)}$ and $N_{(L)}$ is the number of atoms in the cluster when it enters and exits the cell respectively. All droplets below a certain size $N_{min}$ will not make it through the cell and completely evaporate. Therefore, the equation may be rewritten as:

$$\dot{N}_2^+ = I\Omega \left(1 - \int_{N_{min}}^\infty N_{(L)}(\partial_N C)\,\mathrm{d}N / \langle N \rangle \right)$$

The function $N_{(x)}$ needs a specific model for how much a cluster is going to evaporate when it hits the dopant vapor particles.

### 4.6.1.  Pick-Up and Cluster Shrinkage

A helium cluster of size $N$ is encountering dopant particles and also the evaporated helium itself inside the pick-up cell. $N$ decreases due to evaporation of $\delta$ particles evaporated per encountered scattering particle. The number of evaporated particles increases with the kinetic energy $(3/2)k_B T$. The kinetic energy of the beam is here negligible and so is the helium's binding energy, but $\delta_{Na}$ needs to take the large $(Na_k - Na_m)$-binding energy into account.

Even at our strongest doping, the mean free time between pick-ups is about $2.5*10^{-6}$s. At that point, the temperature of the droplets is independent of the initial conditions at the last guest impact and the temperature of the droplets is about 0.5K. The binding energy of a helium atom to the droplet is [Str87] $E_{binding} = k_B(7.15K - 11.3K/N^{1/3})$, thus, the energy that is taken away per evaporated helium atom is $E = E_{binding} + (3/2)k_B T \approx k_B(7.15K + 0.75K) = k_B * 7.9K$ and this results in $\delta_{He} = (3/2)T/7.9K$.

In order to integrate $N' = dN/dx$ along the pick-up path $0 \le x \le L$ one needs to consider $k' = dk/dx$, where $k$ is the number of encountered particles. It holds $N' = -(k'_{Na} \delta_{Na} + k'_{He} \delta_{He})$ and the number of encountered particles grows with $k' = F\sigma n$ as explained in the introduction, that is for example a droplet cross section of $\sigma = BN^a$. So the differential equation reads:

$$N^{-a}dN = -B\sum\nolimits_{i=Na}^{He}(F_i n_i \delta_i)dx$$

If $\delta$ does not depend on k, $\int_0^x dx = \int_{N_{(0)}}^{N_{(x)}} dN$ leads to:

$$N_{(0)}^{1-a} - N_{(x)}^{1-a} = (1-a)Bx\sum(Fn\delta)$$

$N_{(x)} \ge 0$ yields the boundary condition $N_{(0)} \ge N_{min}$ with $N_{min} := [(1-a)BL\sum(Fn\delta)]^{1/(1-a)}$. After defining $x_{max} := N_{(0)}^{1-a}/((1-a)B\sum(Fn\delta))$ follows:

$$N_{(x)} = \begin{cases} [N_{(0)}^{1-a} - xN_{min}^{1-a}/L]^{1/(1-a)} & ; x \le x_{max} \\ 0 & ; x \ge x_{max} \end{cases}$$

Two remarks are in place: Firstly, this folded with a cluster size distribution leads to a new size distribution existing after the pick-up cell. The new distribution needs to be re-normalized because all the clusters that have size $N = 0$ are not counted anymore. Secondly, defining $k_{max} := N_{(0)}/\delta$ results in a symmetric equation for the size $k$ of the guest cluster:

$$\left[1 - \frac{k}{k_{max}}\right]^{1-a} = \left[1 - \frac{x}{x_{max}}\right]$$

Since the droplet cross section decreases, one cannot apply $<k> = \sigma nLF$ anymore. Defining an effective cross section $k_{max} =: \sigma_{eff} n x_{max} F$ results in:

$$\sigma_{eff} = (1-a)\sigma_{(0)}.$$

If $B = \pi r_s^2$ and $a = 2/3$, then $x_{max} = 3N_{(0)}^{1/3}/(\pi r_s^2 \sum (Fn\delta))$, $N_{min} = [\pi r_s^2 L \sum (Fn\delta)/3]^3$,

$N_{(x)} = [N_{(0)}^{1/3} - \dfrac{x}{L} N_{min}^{1/3}]^3$ and $\sigma_{eff} = \sigma_{(0)}/3$.

The surface corrected $B = 4.1\pi r_s^2$ and $a = 13/24$ lead to

$x_{max} = 24N_{(0)}^{11/24}/(11*4.1*\pi r_s^2 \sum (Fn\delta))$, $N_{min} = [11*4.1*\pi r_s^2 L \sum (Fn\delta)/24]^{24/11}$,

$N_{(x)} = [N_{(0)}^{11/24} - \dfrac{x}{L} N_{min}^{11/24}]^{24/11}$ and $\sigma_{eff} = 11\sigma_{(0)}/24$.

### 4.6.2. Total Depletion

Let us concentrate on the host clusters and their atoms, here helium, alone. The steady state is defined by $\dot{N}_2^+ = -\dot{N}_2^-$. The gain on the left hand side was shown to be:

$$I\Omega(1 - \int_{N_{min}}^{\infty} N_{(L)} (\partial_N C) \, dN / \langle N \rangle)$$

Assume the uncorrected shrinkage $N_{(L)} = [N^{1/3} - N_{min}^{1/3}]^3$ and a log-normal size distribution:

$$\partial_N C = \left( N\sqrt{2\pi}(\Delta n) \exp[(\frac{\ln N - \ln \langle N \rangle}{\Delta n} + \frac{\Delta n}{2})^2 / 2] \right)^{-1}$$

Integrating with $N_{min} = \langle N \rangle$ yields:

$$\int_{\langle N \rangle}^{\infty} N_{(L)} (\partial_N C) \, \frac{dN}{\langle N \rangle} = \mathrm{erf}\left[ \frac{\Delta n}{2\sqrt{2}} \right] - 3e^{-((\Delta n)/3)^2} \mathrm{erf}\left[ \frac{\Delta n}{6\sqrt{2}} \right]$$

This amounts to only 0.0048$\langle N \rangle$, meaning that if clusters of average size evaporate completely, less than 0.5% of the clustered helium in the beam will make it through the cell inside of clusters and exit. Even $N_{min} = \langle N \rangle/e$ leads to only 4.3% of clustered helium getting through. What we have shown here is that $N_{min} = \langle N \rangle$/Factor already ensures that practically the whole beam is destroyed. "Factor" could be large depending on how evaporation supplies RMS momentum deflecting the cluster out of the beam and into the cell wall, especially considering the narrow Wittig tube. If basically all clusters evaporate inside the cell, it will hold simply:

$$4I\Omega = a_{(l/d)}A_{\text{tot}}\overline{v}n_2$$

The mean velocity is given via $\left(m_{He}/2\right)\overline{v}^2 = \left(4/\pi\right)k_B T_{cell}$ and with $\Omega = A/R^2$ and $A_{\text{tot}}=2A$ follows:

$$2I = R^2 a_{(l/d)}n_2\sqrt{\frac{8}{\pi}\frac{k_B T}{m_{He}}}$$

Above we showed that $<k> = (1/3)<\sigma_0>n_2 LF$ and that about $\delta_{\text{He}} = (3/2)\text{T}/7.9\text{K}$ helium atoms evaporate per impacting helium atom. Since the whole cluster is depleted, the number of encountered scattering gas atoms is $k = N_0/\delta_{\text{He}}$. These equations plus the cross section formula result in $\left\langle N\right\rangle = \left\langle k\right\rangle\delta_{He} \approx \frac{\pi}{3}r_s^2\left\langle N\right\rangle^{2/3}n_2 LF\delta_{He}$, which will divide the equation dependent on $n_2$ above to yield:

$$I\pi r_s^2\left\langle N\right\rangle^{-1/3}LF\frac{T}{7.9K} = R^2 a_{(l/d)}\sqrt{\frac{8}{\pi}\frac{k_B T}{m_{He}}}$$

## 4.7. Wittig Tube

### 4.7.1. Characteristic Times

The transit time $t_p := Z/<v>$ of the beam through the tube (travel *p*arallel to the beam axis) is about 0.1m over 300m/s [Schî93] (lower *v* and thus longer $t_P$ super-critically), i.e. $t_p \approx 0.3$ms. The time $t_o$ it takes evaporated atoms to reach the wall of the tube (travel *o*rthogonal to the beam) is best evaluated in the center of mass frame of the evaporating droplet. The mean velocity of an evaporated atom is approximated with the mean velocity of particles of a saturated vapor in equilibrium with the evaporating droplet: $\frac{1}{2}m<v>^2 = 4k_B T/\pi$. The over all angles averaged length of the projection of a randomly pointing unit vector $\underline{e}$ is the average $(1/(2\pi)) \int_0^{2\pi} |\sin \alpha| \, d\alpha = 2/\pi$. Thus,

$$v_o = \frac{2}{\pi} \sqrt{\frac{8}{\pi} \frac{k_B T_{\text{drop}}}{m_{He}}} \quad \Rightarrow \quad t_o = r \frac{\pi}{4} \sqrt{\frac{\pi}{2} \frac{m_{He}}{k_B T_{\text{drop}}}}$$

For example, consider a molecule inside $^4\text{He}_{8900}$ absorbing a $18000\text{cm}^{-1}$ photon. The total energy of the droplet is then [Bri90] $16.5\text{K}+25900\text{K}$ and the resulting temperature is $T_{\text{drop}} = 8.69\text{K}$. The atoms evaporated at this temperature inside a $r = 2\text{mm}$ tube will hit the wall after $t_o=15\mu\text{s}$. Even if the evaporation happens at the lower limit of $370\text{mK}$, the time would be also only $t_o = 71\mu\text{s}$.

The evacuation time for helium at r.t. is $\tau = 0.16\text{ms}$ if modeled by $(100, 0, 2)\text{mm}$. The approximation of $a_{(l/d)}$ was only carried out to $0 \leq l \leq 5d$ and the range covered is therefore just $(Z, 0, r)$ to $(Z-20r, 10r, r)$. In the latter limit of $(60, 20, 2)\text{mm}$, the evacuation time is $\tau = 0.5\text{ms}$.

### 4.7.2. At Room Temperature

Between nozzle and entrance aperture is a beam skimmer that hides the tube walls from the source nozzles view in order to avoid parts of the beam striking the tube directly. Thus $R$ is (at least, or larger if the skimmer is smaller than ideal) the distance to the exit aperture (the "stop aperture" in optics terminology). Rearranging the formula for total depletion, the distance $Q := R - (h + 2l)$ to the front aperture follows as:

$$Q = r_s \langle N \rangle^{-1/6} \sqrt{\frac{(h+l)I\pi F \sqrt{\pi m_{He} T_{rt}}}{a_{(l/d)}\sqrt{8 k_B} \, 7.9 K}} - (h+2l)$$

With a $<v> = 270\text{m/s}$ [Sch93] beam and the scattering region at r.t., $F$ for $^4\text{He}$ is about 4.7 already. If $<N> = 50000$ and $I = 10^{20}\text{Hz/sr}$, the $(100, 0, 2)\text{mm}$ model of the Wittig tube requires negative distances from the source, i.e. the steady state of complete beam depletion is impossible. The (60, 20, 2)mm model requires $Q \leq$ 57mm. This would leave no space for a scattering cell.

Using the surface corrected formula from chapter "Theoretical II" results in a large correction. It gives a completely evaporating cluster a bigger cross section when it goes through the small sized end phase of its life. Although the formula $\sigma \approx 4\pi r_s^2 N^{13/24}$ still underestimates the actual cross section, it leads to $Q \leq 90$mm via:

$$Q = r_s \left\langle N \right\rangle^{-11/48} \sqrt{4.1 \frac{11(h+l)I\pi F \sqrt{\pi m_{He} T_{rt}}}{8 a_{(l/d)} \sqrt{8 k_B} 7.9K}} - (h+2l)$$

This result would have been hard to get with published "better" surface corrections that are not engineered to mathematical convenience for the task at hand.

The conditions of $P_0 = 20$bar, $T_0 = 10$K, 5μm nozzle diameter and droplet size of about $<N> \approx 50000$ leads to $I = 10^{19}$Hz/sr of clustered helium atoms [Toe04]. If this case, the steady state of a beam self destructing inside the Wittig tube is impossible if the tube is not heated. That does not imply that there is no enhancement of depletion at all. On the other hand, even if the stable extreme state were on principle possible with the above parameters, it would be still very far from being readily produced by a runaway process. This argues against enhancement of depletion.

### 4.7.3. The Heated Tube

The temperature dependent term $TF^2$ is proportional to the fourth power of the distance $R$:

$$TF^2 = \left\langle N \right\rangle^{2/3} \left(R/r_s\right)^4 \left(\frac{a_{(l/d)} 7.9K}{I(h+l)}\right)^2 \frac{8 k_B}{\pi^3 m_{He}}$$

That $F$ is also $T$-dependent creates the need to solve the equation iteratively:

$$T = \text{Corr} \left\langle N \right\rangle^{\frac{2}{3}} \left(\frac{Q+h+2l}{r_s}\right)^4 \left(\frac{a_{(l/d)} 7.9K}{I(h+l)F}\right)^2 \frac{8 k_B}{\pi^3 m_{He}}$$

"Corr" takes care of corrections:

$$\text{Corr} = \begin{cases} 1 & ; \quad \text{uncorrected} \\ 64 \sqrt[4]{\langle N \rangle} / 45.1^2 & ; \quad \text{surface corrected} \end{cases}$$

Let us put the tube as close as practically possible via $Q = 1$cm. Using a beam speed of 270m/s, $\langle N \rangle = 50000$ and $I = 10^{19}$Hz/sr, the surface corrected formula predicts $T = 1000$K with $F = 8.6$ for the (60, 20, 2)mm model.

## 4.8.  Secondary Helium Assisted Total Depletion in a Scattering Cell?

For the pick-up cell, the pick-up path is $L = 2r + l$ and we must check that in the end $x_{\max} \leq L$. Using $R = 0.5$m, a beam speed of 270m/s, $I = 10^{20}$Hz/sr and the dimensions of our pick-up cell, results in 1575K ($F = 10.7$). The equation is uncorrected though:

$$T_{cell} = \left( \frac{\langle N \rangle^{1/3} 7.9K * R^2 a_{(l/d)}}{I \pi r_s^2 (2r+l) F} \right)^2 \frac{8k_B}{\pi m_{He}}$$

The surface corrected formula

$$T_{cell} = \langle N \rangle^{\frac{11}{12}} \left( \frac{7.9K * (8/11) R^2 a_{(l/d)}}{4.1 * I \pi r_s^2 (2r+l) F} \right)^2 \frac{8k_B}{\pi m_{He}}$$

predicts $T_{cell} = 1078$K with $F = 8.86$. Considering that $I = 10^{19}$Hz/sr may be closer to the truth, secondary Helium atoms seem not to deplete the beam. Using a $k$-dependent $\delta_{Na}$ (e.g. adding 260 evaporated helium atoms for each binding to a trimer $Na_3$) also argues for complete domination by dopant scattering (unpublished spreadsheet). Nevertheless, $I$ is much larger for beams with exp-distributed clusters from strongly super-critical expansions whose velocities are also much slower, for example $T_0 < 4.2$K leads to very large droplets $\langle N \rangle \approx 10^{10}$ with velocity of only 15 m/s [Gri03]). Also consider that our cell is unusually far from the source ($R = 0.5$m) and $T \sim R^4$. For compacter setups and high intensities $I$, secondary Helium atoms inside the pick-up cells may become an issue that complicates beam depletion calculations.

### 4.9. Evaluation and Outlook

The model above neglects that evaporating clusters amass rms−momentum orthogonal to the beam axis. If clusters hit the cell or tube walls, they will completely evaporate. Since the Wittig tube is long and narrow, this could be the deciding factor and make the idea workable in the strongly super critical regime where intensities are higher and beam velocities much below the above assumed 270m/s. A pulsed nozzle [Sli02] may not help. It has instantaneous intensities of $I = 10^{22}$Hz/sr, but only for very short pulses of about 1cm length.

Calculating $\delta_{He}$ at the cluster temperature of 2K overestimates depletion, because if the clusters are actually hotter than this, an evaporated atom will take more energy away and fewer evaporated atoms are necessary to carry the energy. Moreover, especially the small clusters have not enough total energy to evaporate atoms efficiently. Here a correction could be put relatively fast with the same technique as done to correct for the low density of the droplets surface layer.

We neglected all the uncondensed helium monomers that travel inside the beam. They have small cross sections but are numerous. A single collision with a dopant atom or another helium particle inside the cell/tube will add such a monomer immediately to the thermalized scattering gas.

Lastly, the evacuation times match usual chopping periods and common pulsed laser repetition times. This leads to interesting ideas of how to match the chopping frequency to the cell or tube and what the variation of the chopping frequency may be able to probe.

# 5        Theoretical II: Size Distributions

## 5.1.  Statistical Distributions

Processes that randomize the results of other random processes further give rise to typical size distributions. Among these are normal, log-normal (LN) and exponential (EXP). While decay or fractionation leads often to power laws or EXP distributions, random grow processes like phase change aggregations lead often to normal or LN results, be it in biology, economics or cluster physics [Vil93, Wan94].

Mean and deviation are independent degrees of freedom (DOF) of a LN but they often seem proportional to each other. For beams of $He_N$, $<N> \approx 1.1*FWHM$ (full width at half max) is established [Har98] and we have seen in the introduction why the three DOF ($P_0$, $T_0$, and nozzle size) reduce to one continuous parameter.

For the EXP distribution $<N> = \Delta N$ holds exactly, although growth seems to be quite the opposite to violent destruction processes. Cluster beams can in a sense connect these opposites smoothly. Sub (*super*) critical expansions produce condensation (*fractionation*) clusters that are distributed log-normally (*exponentially*) in size. Apart from the supercritical expansion also being bimodal, the expansion is able to connect statistical cluster growth and statistical decay into clusters.

For clusters produced by an expansion, the cluster size probability distribution only cares about where the expansion trajectory through phase space intersects the bi-nodal and from which side it does so. While climbing the bi-nodal along the vapor side, $<N>$ increases and continues to increase when turning around at the critical point and descending along the liquid side of the coexistence. Can the distribution obtained when intersecting from one side of the coexistence be mapped into the one when intersecting from the other side? One may argue against this: Certainly, in a Mollier diagram (enthalpy versus entropy) or a pressure versus density diagram, the coexistence line opens up into an 2D area of expansion conditions and the intersections that seem on top of each other in the $P$ versus $T$ plot are widely apart.

### 5.1.1. General Probability Distributions

Consider a statistical variable $m$ in the real numbers with min $\leq m \leq$ max and mean denoted as $<m>$. The standard deviation $\Delta m$ is defined via $(\Delta m)^2 := <m^2> - <m>^2$. Any linear transformation $m := an + b$ leaves its normalized variable $\tilde{n} := (m - <m>)/\Delta m$ untouched: $n = \tilde{n}\Delta n + <n>$, i.e. $a = \Delta m/\Delta n$. It helps knowing that $n = \ln N$ will refer to number of particles $N$ and $m = \ln M$ often to the droplet cross section $M$. The following is general though and useful also for the exponential distributions.

A succinct, powerful origin for a probability distribution is the cumulative probability $C$. Cumulative means that $C = \int_0^{C_{(m)}} dC = \int_{\min}^{m} (dC/dm)\, dm$ , i.e. the infinitesimal probability of any $m$ is d$C$. Once such a function is equal to unity at $m =$ max, it leads to automatically normalized distributions that hold for all $m = an+b$ if it is written as $C_{(\tilde{n})}$. Its expectation values are $\langle \Psi \rangle = \int_0^1 \Psi dC$ which, if boundaries are at infinity for example, equals $\langle \Psi \rangle = \int_{-\infty}^{+\infty} \Psi (dC/d\tilde{m})\, d\tilde{m}$. With $(d\tilde{m}/dm) = 1/(\Delta m)$, the most likely value of $m$ (modal value) is at $(d^2C/dm^2) = 0$ if a maximum exists away from the boundaries at min and max. With $b = \ln B$ follows $m = \ln(BN^a)$.

The distributions follow as $(dC/dM)$. The definition implies $dM = M\, dm$ and the distributions for $M$ are therefore equal to the ones for $m$ yet divided by $M$:

$$(dC/dM) = M^{-1}(dC/dm).$$

The expectation of $\Psi$ is thus written:

$$\langle \Psi \rangle = \int_0^\infty \Psi (dC/dM)\, dM = \int_0^\infty (dC/dm)(\Psi/M)\, dM$$

### 5.1.2. Normal and Lognormal Distribution

The cumulative normal probability is $C_n := \frac{1}{2}\left[1 + \mathrm{erf}\left(\tilde{n}/\sqrt{2}\right)\right]$. The normal distribution of $m$ follows as $dC_n/d\tilde{m} = [\sqrt{2\pi}]^{-1}\exp\left(-\tilde{m}^2/2\right)$. The modal value equals the mean $<m>$. The log-normal (LN) distribution follows as $\left(dC_n/dM\right)$ and is again just the one for $m$ but divided by $M$.

The derived LN expressions are not convenient. The p$^{th}$ moment $<M^p>$ turns out to be involved and may be expanded via $M = e^m$ and $e^m = \sum_{i=0}^{\infty}\left(m^i/i!\right)$. It holds $\left\langle M^p \right\rangle = \exp\left[p\left\langle m\right\rangle + (1/2)\left(p\Delta m\right)^2\right]$. Hence, the modal $\exp[<m> - (\Delta m)^2]$ occurs before the mean $<M>$, which now always transforms with a shift: $\left\langle M^p \right\rangle = \left\langle M\right\rangle^p \exp\left[\left(p/2\right)(p-1)\left(\Delta m\right)^2\right]$.

The length $2\Delta N$ centered at the modal of d$C$/d$N$ likely reaches back below $N < 0$, thus the FWHM is often used. A general deviation FWXM =: $(\Delta M)_X$ is centered at the modal of $N$. $(\Delta M)_X = (E_+ - E_-)$, with $\ln[E_\pm] = <n> - (\Delta n)^2 \pm (\Delta n)\sqrt{(-2\ln X)}$. From the normal distribution's point of view, all this is unnecessary. $M < 0$ is excluded because no deviation stretches below $m = -\infty$.

To focus onto $\ln(M)$ can be advantageous if $M$ varies over several orders of magnitude and has an absolute zero, like temperature. If $M$ distributes log-normally, $m$ = $\ln(M)$ will be as meaningful even if it describes something physically quantized into integers like particle number. The LN is a continuous distribution with no convenient binning to render it discrete. Employing the LN means already having given up $M$'s integral nature.

### 5.1.3. Cluster-physical Distributions: The Sub-critical

For helium droplets, data have been fitted with LN distributions and fitting parameters $<n>$ and $\Delta n$. Originally, $<N>$ and $(\Delta N)_{1/2}$ were calculated with $<N>$ = $\exp(<n> + \frac{1}{2}(\Delta n)^2)$ and $(\Delta N)_{1/2} = E_+ - E_-$, where the choice $X = 1/2$ determines the proportionality factor $-2\ln X$. With these choices it follows the proportionality $<N>/(\Delta N)_{1/2} \approx 1.1$ from the data.

However, notice that $<N>/(\Delta N)_X = \exp(<n> + \frac{1}{2}(\Delta n)^2)/(E_+ - E_-)$ = $\exp[(3/2)(\Delta n)^2]/\{\exp[(\Delta n)\sqrt{(-2\ln X)}] - \exp[-(\Delta n)\sqrt{(-2\ln X)}]\}$, i.e., $<n>$ is actually not inside this equation. The proportionality depends strongly on the measure FW$X$M (there is nothing fundamental about $X = \frac{1}{2}$). $X = 0.438$ would lead to the proportionality being equal to unity: $<N> = (\Delta N)_X$. Moreover, once $<n>$ is removed from the procedure, one realizes that the proportionality varies weakly with $\Delta n$. In fact, the surprising relation $<N>/(\Delta N)_{1/2} \approx 1.1$ would be equally true after setting all the data for $\Delta n$ to a constant around 0.6. Why is this so? The origin of the surprisingly restrictive proportionality $<N> \sim \Delta N$ is hidden by the usage of $<n>$ and $\Delta n$ as fitting parameters and the subsequent transformation (the move from $n$- to $N$-space) into two variables that hardly have any dependence on $\Delta n$; e.g. $<N>$ depends obviously mostly on $<n>$, but so does $(\Delta N)_{1/2}$. The whole sub-critical range explored extends over a range of only $<n> = 7 \pm 3$. In this sub-critical range of all the helium experiments done to date holds $\Delta n = 0.55 \pm 0.15$.

For smaller droplets, the low density of the droplets' surface increases the geometrical cross section much over the simple liquid drop model $\sigma_{\text{l.d.}}$. This suggests that cross sections are not LN distributed, yet, given the accuracy of experiments, $N$, $\sigma$, both or even neither deviate from a LN distribution. [Von10] argues for using the Inverse Gaussian (IG) distribution.

### 5.1.4. Unification with the Supercritical Expansion's Distribution

In the supercritical regime, liquid helium fragments into droplets; the size distributions fall off exponentially at large $N$ [Knu99]. However, using the linear EXP for $N$ gives it a special status: If $N$ is EXP distributed, $M$ is not. Using the regularized gamma ($\Gamma$) function as the cumulative $C_g := 1 - \Gamma[d, (dN/<N>)]/\Gamma[d]$, the $\Gamma$-distribution is derived. For $d = 1$ it yields the EXP distribution, but it always has the exponential fall off that is observed at high arguments $N$. Its deviation is $\Delta N = <N>/\sqrt{d}$. For $d > 1$ there is a modal and the $\Gamma$ function looks like a LN in that case.

Consider the cumulative probability $C_{EXP} := 1 - \exp(-N/<N>)$. The EXP distribution of $M$ follows as

$dC_{EXP}/dM = \exp(-N/<N>) \, (d(N/<N>)/dM) = \exp(-N/<N>) \, N/(<N>M\Delta m)$. We yield $dC_{EXP}/dN = <N>^{-1} \exp(-N/<N>)$ and this is again just dividing $n$'s distribution by $N$, therefore $dC_{EXP}/dn = <N>^{-1} \exp[n - <N>^{-1} \exp(n)]$. This is the "exp-exponential" distribution to be consistent with the usual "log" that is added to "normal". While a monotonic exponential decline cannot have a modal, in $n$-space the modal equals $\ln<N>$. It holds furthermore $<N> = \exp(<n> + \gamma)$ where $\gamma$ is the Euler-Mascheroni constant:

$$\gamma = \lim_{G \to \infty}\left[\ln\left(1/G\right) + \sum_{g=1}^{G}\left(1/g\right)\right] \approx 0.5772$$

Thus $dC_{EXP}/dn = \exp[n - <n> - \gamma - \exp(n - <n> - \gamma)]$, which looks less weird than the LN expressed without $n$, $<n>$ and $\Delta n$. $\Delta N = <N>$ turns into $\Delta n = \pi/\sqrt{6}$, i.e. it is exactly fixed, reminiscent of the discussion involving the LN. We could deduce again that the fixed dispersion may rather be in the assumption of a certain distribution used to fit the data rather than an experimental result.

The $\Gamma$, IG, Generalized IG, or the Power IG may unify the expansion regimes smoothly with less DOF than the LN. This should be done while trying to relate to other work that expresses size distributions' variables dependent on physical properties that depend on the dimensionality of spaces involved [Vil93, Knu97]. Leaving corrections to $\sigma_{l.d.} = \pi \; r_s^2 \; N^{2/3}$ for later, consider the ansatz $\Delta n \propto \Delta(\ln \sigma_{l.d.}) \propto (\text{dim}-1)/D$. The one dimension of the beam's axis is subtracted from the dimensionality dim of the space into which the expansion occurs. Then it is divided by the full spatial dimensionality $D$. (dim$-1$) is usually (3$-1$) = 2 as above but a much narrower distribution would be predicted due to (2−1) = 1 in case of a long slit aperture. This suggests that the size distribution can be sharpened by using a long slit instead of a circular aperture, effectively reducing dim. The limits are immediately meaningful: (dim−1) = (1−1) = 0 would constitute the consistent extreme $\Delta m = 0$ if there is no expansion and therefore no clustering in case of an attached infinitely long tube instead of a nozzle or say an infinitely huge "aperture".

## 5.2. Surface Corrections and Fractal Dimension

One may derive other distributions with desired mathematical forms that include surface corrections for example, as we have used several times before in this thesis. This procedure may be useful in general cluster physics. The aim here was to engineer expressions that combine with other probabilities for beam depletion, pick-up and so on in a way that leaves the folding analytical.

First Example: The ansatz $\sigma =: \sigma_{l.d.} + M'$ makes combining the expectation values trivial: $<\sigma> = <\sigma_{l.d.}> + <M'>$. Taking experimental data [Har98] for $<\sigma>$ and fitting $\ln[<\sigma> − <\sigma_{l.d.}>]$ versus $\ln<N>$ results in the pair $(a', B')$. The derived $M' \approx \pi(9/2)r_s{}^2 N^{4/11}$ describes all data within experimental accuracy. The correction is not an added surface layer but rather redistributes the atoms of the liquid drop model giving them less density to get the right cross section $\sigma_{(N)}$ for any *single* droplet $N$, all by virtue of the normal statistics of a certain very specific statistical ensemble. (Starting with a *single* drop $He_N$ instead, one may divide the drop into a core of radius $r_{(N)}$ having bulk density and a surface layer having $N_{surface}$ proportional to $4\pi(r+2\text{Å})^2$ with a surface-density-constant as the proportionality factor. Adjusting the involved constants using the same data as training set, the agreement between experiment and this for a single droplet "physically justified" model ($\sigma = \sigma_{core} + \sigma_{surface}$) is no better.)

Second Example: The ansatz $\sigma =: \sigma_{l.d.}M'$ leaves $\sigma$ and $N$ log-normal and is equivalent to a simple fitting of $<\sigma>$ via linear regression ($m = a\,n + b$). It results in $M' \approx 4.1*N^{1/8}$, i.e. $\sigma \approx 4.1\pi r_s{}^2 N^{13/24}$ and $r \approx 2r_s N^{13/48}$, which translates to a fractal dimension of $^4$He clusters of $D = 48/13 \approx 3.7$ in terms used by [Vil93]. This ansatz has been used in the section on the "Wittig Tube" and its convenient structure led to a very fast but impressive correction there, although it underestimates cross sections of small clusters. The ansatz could be worked out to correct for the slow evaporation of small clusters, always keeping analyticity of the solutions.

# 6      Theoretical III: Spin Desorption Statistics

Observations on alkali atoms agglomerating have been interpreted with the suggestion [Sch04] of novel quantum systems of fermions bound together weakly (vdW). All fermions are completely spin polarized resulting in clusters with giant total spins swimming on top of sub-Kelvin helium. The reasoning goes as follows:

The energy liberated upon covalent condensation of for example $Na_4$ is sufficient to eject the particle from its position on the surface dimple of the droplet. Once desorption occurs, a droplet starts over again to pick up guests along the rest of the path $L$. This selects for fully spin polarized clusters, because they are only vdW bound. The violent interaction with the detection equipment causes spins to flip such that the cluster suddenly binds covalently. The mass spectra reflect the products of this binding, i.e. usual alkali clusters $A_k$, yet dissociated to smaller sizes due to the large binding energy release. It has been discovered [Sti95, Sti96, Hig96, Hig98] that high-spin states (i.e., $Na_2$ triplets and $Na_3$ quartets) are produced efficiently by using the pickup process. These states remaining intact on $He_N$ surfaces for long times. They are bound to the helium cluster with binding energies of 20 and 100meV, respectively [Hig96]. Such amounts of energy can be dissipated readily by the evaporation of less than 12 and 60 He atoms, respectively. It is noteworthy that the $Na_2$ and $Na_3$ ground states are significantly less populated than the high-spin states (e.g., by up to three orders of magnitude in the case of $Na_2$) [Hig96]. This is consistent with facile sodium desorption from the cluster due to the energy of 0.7 to 1eV per atom [Dug97] that is released upon forming the ground electronic states.

A more conservative approach would argue that desorption upon covalently binding one more monomer into a cluster is not as efficient as the super fluid matrix is in carrying away binding energy. Clusters do desorb, but a portion stays even through the bottle neck at $k = 4$, after which the clusters become larger and less likely to desorb. The probability of many spins $s = \pm 1/2$ to come together randomly as fully spin aligned must go down by a factor of $2^k$ and is very unlikely indeed for the large $k \approx 20$ to 30 observed. Moreover, for larger clusters, it is unrealistic to expect a comparable degree of stability for high-spin states [Med95]. The mass spectra reflect the abundance of low spin clusters that are dissociated by a lot less than their whole covalent binding energy. The ionization potential for alkali is very small and dissociation due to excess energy is very likely.

Note that both views can qualitatively account for the odd-even stability pattern observed. [Von03] investigated the final size distribution of clusters consisting of $k$ spin $s = \frac{1}{2}$ particles combining to total spin $S$, that aggregated with the help of helium droplets. Incident (initial) droplet size distributions and desorption statistics depending on spin alignment were considered. To make reviewers happy, a wrong derivation had to be given. This is a good opportunity to provide the proper derivation of the main, unadulterated result. This is well in place, because it calls into question part of our reasoning above and in [Von02]. Starting from this, a more careful treatment should be started once more again.

## 6.1. Without Desorption

A helium droplet picks up $k$ guest atoms that are spin doublets $\underline{2}$ with spin $s = 1/2$. The guests aggregate to clusters and the spin is conserved. For the overall spin multiplet holds $\underline{2} \otimes \underline{2} \otimes \ldots = \underline{2} \otimes \underline{2}^{k-1} = \underline{2}^k$. For example, having picked up $k = 0$ particles leads to the singlet $\underline{2}^0 = \underline{1}$. For all $d \geq 2$ holds $\underline{2} \otimes \underline{d} = \underline{d+1} \oplus \underline{d-1}$, where $d = 2S+1$ is the degeneracy. This suggests to set up a recursion relation via the definition of expansion

parameters $(n, k)$:

$$\underline{2}^k =: \underline{k+1} \oplus \sum_{n=0}^{n \leq (k-2)/2} (n,k) \underline{d}$$

The degeneracy is $d = (k-2n-1) = 2S+1$ and the recursion relation $(n+1, k+1) = (n, k) + (n+1, k)$, where $n$ is an integer $n \geq -1$, i.e. $\underline{k+1}$ is the $n = -1$ multiplet and $S$ is the total spin. Enumerating the $(n, k)$ starts with $(0, k) = (k-1)$ and $(1, k) = \frac{1}{2}(k-3)\,k$. Complete induction gives:

$$(n,k) = \frac{d}{(n+1)!} \prod_{q=0}^{n-1} (k-q) = \frac{d}{(n+1)!} \frac{k!}{(k-n)!} = \frac{d}{n+1} \binom{k}{n}$$

The definition $a_\pm := \frac{1}{2}\,[k \pm (d+1)]$ leads to $(n, k) = k! d / [a_+! \; a_-!]$ and reveals the origin of these formulas in the theory of Young tableaux. The number of states equals $\sum_{n=-1}^{n_{max}} (n,k)\, d$ and since the alignment of spins of picked up atoms is random, the number # of states in the statistical ensemble is: $\# := (k+1) + \sum_{n=0}^{2n \leq k-2} (n,k)\, d$ , and this is equal to the number of states of $\underline{2}^k$, as it must:

$$\# = k+1 + \sum_{n=0}^{2n \leq k-2} \frac{d^2}{n+1} \binom{k}{n} = 2^k$$

All clusters are $d$-lets, that is multiplets $\underline{d}$ with multiplicity $d = (k-2n-1)$. Indexing with $<k>$ by writing $^{<k>}P$ means that the probability is dependent on a statistically distributed k with expectation $<k>$. Therefore, writing $^k P$ stresses that we still are discussing the consideration that a certain number $k$ of guests is picked up. In other words, $<k> \equiv k$, or as far as $^k P$ is concerned, it does not know about other values of $k$ being possible. The sum of all probabilities $^k P_{\underline{d}}$ of all $d$-lets has to be normalized with $\sum_{n=-1}^{n_{max}} {}^k P_{\underline{d}} = 1$. Using Kronecka deltas $^n \delta_m = \delta_{n,m}$, the result is written as:

$$^k P_{\underline{d}} = 2^{-k} d \left[ \frac{d}{n+1} \binom{k}{n} \right]^{1-\delta_{n,-1}}$$

Our interest lies in the probability of cluster masses, not only spins. There is no desorption yet, so all clusters are $k$-mers: $^k P_{l-\mathrm{mer}} = {}^k \delta_l$. The aim of course is to deal with distributions $^{<k>}P_k$ of $k$. Observables are modeled by expectation values. For example, the probability to observe an $l$–mer is $P_{l-\text{mer}} = \Sigma_{k=0}^{\infty} {}^{<k>}P_k {}^{\,k}P_{l-\text{mer}}$ and normalization $\Sigma_{k=0}^{\infty} {}^{<k>}P_k = 1$ leads to $P_{l-\text{mer}} = \Sigma_{k=0}^{\infty} {}^{<k>}P_k {}^{\,k}\delta_l = {}^{<k>}P_l$. Indexing in this way assures consistent notation, i.e. notation-wise nothing changes between considering certain fixed $k$ or distributions of $k$. Secondly, one may adopt the Einstein convention to greatly facilitate deriving: whenever a lower index $k$ and an upper one meet, the sum over all possible $k$ is implied and indices are "contracted". To clarify usage: I just wrote "the probability to observe an $l$–mer …". That derivation shortens now to $P_{l-\text{mer}} = {}^{<k>}P_k {}^{\,k}P_{l-\text{mer}} = {}^{<k>}P_k {}^{\,k}\delta_l = {}^{<k>}P_l$. Note that the power of the approach rises and falls as it does for contra and covariant coordinates or the bras $<|$ and kets $|>$ of quantum physics. For example, normalization is written $<1\ |> = <1> = 1$ while $|>1<|$ is an operator and "something entirely different". Here, the observable expectation, that is the guest cluster size (or spin) distribution $P_{l-\text{mer}}$, equals the pick-up statistics $^{<k>}P_k$ times the desorption statistics $^{k}P_{l-\text{mer}}$.

If the pick-up statistics is Poissonian ($^{<k>}P_k = \text{e}^{-<k>}<k>^k/k!$), then so is the distribution of guest cluster sizes: $P_{l-\text{mer}} = {}^{<k>}P_l$ ; $l \in \mathbb{N}^0$.

### 6.2. Desorption of Spin Relaxing Clusters

Under the assumption that any two guest atoms bind covalently if not spin aligned and assuming desorption of the cluster from the droplet upon the release of covalent binding energy, the $\underline{2} \otimes \underline{d} = \underline{d+1} \oplus \underline{d-1}$ above goes to $\underline{2} \otimes \underline{d} = \underline{d+1} \oplus (d-1)\underline{1}$, again for all $d \geq 2$. Once a helium droplet desorbs a cluster, it starts over again and picks up more guest atoms, therefore $\underline{2}^k = \Sigma_{n=-1}^{k} 2^{|n|-1} \underline{d}(1-\delta_{n,0})$. The degeneracy is here $d = (k-n) = 2S+1$. The number of states in the statistical ensemble is $\Sigma_{n=-1}^{k} 2^{|n|-1}(k-n)(1-\delta_{n,0}) = 2^k$. All multiplets $\underline{d}$ are now $(k-n)$-lets and normalization $\Sigma_{n=-1}^{k} {}^{k}P_{\underline{d}} = 1$ yields $^{k}P_d = d2^{|n|-1-k}(1-\delta_{n,0})$.

The latter assumes that spins do not interact with the droplet surface; i.e. it does not "measure" spin according to its normal vector defining a $z$-axis or at least spins are not redirected - say with a preference for large $|S_z|$ and bosonic bits of angular momentum taken up by the helium.

The clusters are now $(k-n-1)$-mers. Considering again that $\sum_{l=0}^{\infty} {}^k P_{l-\text{mer}} = 1$ and ${}^k P_{(k-n-1)-\text{mer}} = {}^k P_{d/\underline{d}}$ results this time in:

$$^k P_{l-mer} = (l+1)2^{|k-l-1|-k-1}(1-\delta_{k-1,1}) \quad ; l \in \{0,1,...,k\}$$

Interestingly, this alone is already quite different from what one would expect a mass spectrum to look like. The numbers are here mostly independent of $k$:

| $l$ | 0 | 1 | 2 | 3 | 4 | 5 | ... | $k$-1 | $k = <k>$ |
|---|---|---|---|---|---|---|---|---|---|
| ${}^k P_{l-\text{mer}}$ | 1/4 | 1/4 | 0.188 | 1/8 | 0.078 | 0.047 | ... | 0 | $(k+1)2^{-k}$ |

and $\langle l \rangle = \sum_{l=0}^{k} l \left( {}^k P_{l-mer} \right) = 2(1-2^{-k})$.

This feature does not change much with a different pick-up statistics ${}^{<k>}P_k$ with allowed $k \neq <k>$, because practically one needs very high $<k>$ and physical probability distributions are much different from ${}^{<k>}P_k \approx 0$ only around the expectation $<k>$. Detecting a finite few magnitudes of signal strength means then to observe the tabulated factors. Constant ratios of mass peaks have been observed and explicitly related [Von02] to fractionation upon detection rather than spin statistics!

Applying $P_{l-\text{mer}} = {}^{<k>}P_k{}^k P_{l-\text{mer}}$ like above yields now a much more complicated dependence:

$$P_{l-mer} = (l+1)\sum_{k=l}^{\infty} {}^{<k>}P_k 2^{-k}(1-\delta_{k-1,1})$$

For example, if the pick-up statistics is Poissonian (${}^{<k>}P_k = e^{-<k>}<k>^k/k!$), it follows:

$$P_{l-mer} = {}^{<k>}P_l \frac{l+1}{2^l} + \left( l+1 - \frac{\Gamma_{(l+2,<k>/2)}}{l!} \right) e^{-<k>/2}$$